\documentclass[11pt]{article}

\pdfoutput=1

\usepackage{times}
\usepackage{fullpage}
\usepackage{graphicx}
\usepackage{bm}
\usepackage{amsfonts}
\usepackage{amssymb}
\usepackage{amsmath}
\usepackage{amsthm}
\usepackage{subcaption}
\usepackage{appendix}
\usepackage{booktabs}
\usepackage{siunitx}

\title{Effective transient behaviour of inclusions in diffusion problems}

\author{L. Brassart\\
		Department of Materials Science and Engineering\\
		Monash University\\
	     Clayton, VIC~3800, Australia\\[.5em]
	    \and L. Stainier \\  
	    Institut de Recherche en G\'enie Civil et M\'ecanique\\
	    (GeM, UMR 6183 CNRS-ECN-UN)\\
	    {\'E}cole Centrale Nantes, F-44321 Nantes, France 
	}

\begin{document}
\maketitle

\newtheorem{remark}{Remark}

\begin{abstract}
This paper is concerned with the effective transport properties of heterogeneous media in which there is a high contrast between the phase diffusivities. In this case the transient response of the slow phase induces a memory effect at the macroscopic scale, which needs to be included in a macroscopic continuum description. This paper focuses on the slow phase, which we take as a dispersion of inclusions of arbitrary shape. We revisit the linear diffusion problem in such inclusions in order to identify the structure of the effective (average) inclusion response to a chemical load applied on the inclusion boundary. We identify a chemical creep function (similar to the creep function of viscoelasticity), from which we construct estimates with a reduced number of relaxation modes. The proposed estimates admit an equivalent representation based on a finite number of internal variables. These estimates allow us to predict the average inclusion response under arbitrary time-varying boundary conditions at very low computational cost. A heuristic generalisation to concentration-dependent diffusion coefficient is also presented. The proposed estimates for the effective transient response of an inclusion can serve as a building block for the formulation of multi-inclusion homogenisation schemes.\\


\end{abstract}

\newpage

\section{Introduction}
The general context of this work is the identification of macroscopic continuum theories to describe diffusive transport in heterogeneous media. This is a classical homogenisation problem, and a number of analysis methods have been proposed, including semi-analytical bounds and estimates \cite{hashin1968,benveniste1986}, asymptotic homogenization theory \cite{auriault1983,auriault2009} or volume-averaging methods \cite{quintard1993,moyne1997,whitaker1999}. These approaches usually require the solution of a boundary value problem defined on a Representative Volume Element (RVE) of the microstructure in order to identify effective parameters to be used at the macroscale. Under the separation of scales hypothesis, one often assumes that the microfields instantaneously reach a steady-state in the RVE, while transient diffusion is handled at the level of the macroscopic boundary value problem. 

The assumption of microscale steady-state becomes however questionable when the length scale of the macroscopic excitation becomes comparable to the size of the RVE, or when there is a high contrast between the diffusivities of the phases. In the latter case, transport through the slow phase induces a memory effect at the macroscopic scale. Microscopic transient effects are particularly relevant for mass diffusion problems, as compared to heat conduction problems, as the diffusion coefficients between the phases can vary by orders of magnitude \cite{moyne1997}. 

Memory effects due to contrasted transport properties were formally analysed by Auriault \cite{auriault1983} using asymptotic homogenisation applied to heat transfer problems. He showed that the memory effect due to transport in the slow phase translates into a macroscopic heat conduction equation involving the convolution integral of a memory function, to be identified from a transient analysis on the slow phase. The author further provided some analytical expressions for the memory functions on simple geometries, including laminate composites, and cylindrical and spherical inclusions. The analysis was later adapted to mass transfer \cite{auriault1995}. Recently, Dureisseix et al. \cite{dureisseix2015} developed a finite element (FE) approach to evaluate the incremental evolution of the memory function for periodic microstructures by solving the transient diffusion problem in the slow phase, and also proposed heuristic semi-analytical estimates.  An asymptotic homogenisation approach was also developed by Matine et al. \cite{matine2015} to include edge and short time effects in transient heat conduction in laminate microstructures.

In the present work we address the memory effect in two-phase composites consisting of "slow" inclusions dispersed in a "fast" percolating matrix. Our goal is to develop semi-analytical estimates of the transient inclusion response, which can be integrated into a scale transition method aiming at developing macroscopic continuum equations. As such, our results can be combined with different approaches, including asymptotic homogenisation or mean-field schemes. The formulation of a complete homogenization framework for the composite based on the estimates developed in this paper will be addressed in a separate publication. While our proposed approach is formulated in the context of diffusive mass transfer, our results can readily be transposed to heat conduction problems. 

The proposed estimates for the transient inclusion response in linear diffusion problems rely on a formal analogy between the effective diffusive response of an inclusion and the mechanical response of a viscoelastic solid. A chemical creep function is introduced, which relates the average concentration of the inclusion to the applied step load on the inclusion boundary (Section \ref{sec:exact}). The creep function is written as infinite Prony series, and exact expressions of the coefficients in the Prony series are given for simple geometries. We show that knowledge of the chemical creep function enables the determination of the effective inclusion response under a time-varying loading history, similar to the Boltzmann superposition principle of viscoelasticity. 
 
Next we propose systematic approaches to obtain semi-analytical estimates of the chemical creep function for inclusions of arbitrary shape, where an exact expression of the creep function cannot be obtained analytically (Section \ref{sec:estimates}). We propose an efficient numerical time integration algorithm based on a finite number of internal variables, enabling the simulation of arbitrary loading histories at very low computational cost. We show that good estimates can be obtained with a limited number of internal variables (Section \ref{sec:results}). Finally, we propose a simple mean-field approach to extend our results to the case where the diffusion coefficient varies with concentration (Section \ref{sec:nonlinear}).

\section{Position of the problem} \label{sec:problem}
Consider the transient diffusion problem in a closed domain $\Omega$ (an "inclusion"). The medium is assumed homogeneous and isotropic. Let $c(\bm x,t)$ be the local concentration of mobile species (number of molecules per unit volume). Conservation of the number of molecules requires that:
\begin{equation} \label{mass_conservation}
\frac{\partial c}{\partial t} = - \bm{\nabla}\cdot \bm j,
\end{equation}
where $\bm j$ is the diffusion flux. We describe diffusion using Fick's first law:  
\begin{equation} \label{Fick_1stlaw}
\bm j = - D(c)\bm{\nabla} c,
\end{equation} 
where the diffusion coefficient $D$ in general depends on concentration. 
A combination of equations (\ref{Fick_1stlaw}) and (\ref{mass_conservation}) gives the classical Fick second law governing the evolution of the concentration field in space and time:
\begin{equation}\label{eq:Fick_2ndlaw}
\frac{\partial c}{\partial t} = \bm{\nabla}\cdot \left(  D(c) \bm{\nabla} c \right). 
\end{equation}
The inclusion is subjected to homogeneous Dirichlet boundary conditions $c(\bm x,t)=\bar c(t)$ on its boundary $\partial \Omega$. The concentration field at $t=0$ is assumed homogeneous, $c(\bm x,0)=c_0$, and we set $c_0 = 0$ without loss of generality. The concentration $c(\bm x,t)$ can then be viewed as a deviation relative to the initial concentration. In the case where the diffusivity is constant, Equation (\ref{eq:Fick_2ndlaw}) reduces to
\begin{equation}\label{eq:Fick2}
\frac{\partial c}{\partial t} =   D  \bm{\nabla}^2 c. 
\end{equation}

In the search for macroscopic continuum equations through homogenisation, we are mainly interested in the effective inclusion response in terms of its volume average:  
\begin{equation}\label{eq:vol_av}
\langle c \rangle(t) = \frac{1}{V(\Omega)}\int_{\Omega} c (\bm x,t) \ dV,
\end{equation}
where $V(\Omega)$ is the volume of the inclusion. Specifically, we are looking for a mean-field, semi-analytical expression: 
\begin{equation} \label{eq:eff_rel}
\frac{ d\langle c \rangle(t)}{dt} = \mathcal F (\bar c, \langle c \rangle, \bm b),
\end{equation}
where $\bm b$ collectively represents a finite set of internal variables, yet to be identified, describing the effect of past loading history on the instantaneous inclusion response. 

\begin{remark}
In a thermodynamically-consistent formulation of the diffusion problem, the driving force conjugated to the diffusion flux is the negative of the chemical potential gradient, $-\bm{\nabla} \mu$. The kinetic model of diffusion is usually written as 
\begin{equation}\label{diff_kinetics}
\bm j = - cM \bm{\nabla}\mu,
\end{equation}
where $M$ is the mobility, defined as the ratio of the drift velocity of mobile atoms, $\bm j/c$, to the driving force. The chemical potential is generally written as
\begin{equation} \label{chempot_gen}
\mu = \mu_{\textnormal{ref}} + kT \log(a),
\end{equation}
with $\mu_{\textnormal{ref}}$ a reference chemical potential and $a$ the activity of the mobile species, which depends on composition.  A commonly adopted simplified expression of the chemical potential is: 
\begin{equation} \label{chempot_approx}
\mu = \mu_{\textnormal{ref}} + kT \log\left( \frac{c}{c_{\textnormal{ref}}} \right),
\end{equation}
where $c_{\textnormal{ref}}$ is a reference concentration. Expression (\ref{chempot_approx}) is a reasonable approximation for a dilute solution of interstitial atoms in a host \cite{balluffi2005}. If we further assume that the mobility is independent of composition, Expression (\ref{chempot_approx}) is equivalent to assuming a constant diffusion coefficient in (\ref{Fick_1stlaw}). This is easily verified by inserting (\ref{chempot_approx}) into (\ref{diff_kinetics}):
\begin{equation} \label{diff_kinetics_dilute}
\bm j = - MkT \bm{\nabla} c. 
\end{equation}
Comparing (\ref{diff_kinetics_dilute}) with (\ref{Fick_1stlaw}), one recovers the Einstein relation: $D = MkT$. 
\end{remark}

\begin{remark}
In principle, one should consider chemical potential boundary conditions, rather than concentration boundary conditions on the external surface of the inclusions. Indeed, the chemical potential is continuous across the interface, while the concentration is generally discontinuous, and the concentration on the inclusion side is a priori unknown. For a given history of applied chemical potential $\bar{\mu}(t)$ on $\partial \Omega$, it is however possible to invert relation (\ref{chempot_gen}) and obtain the corresponding history $\bar c(t)$.  We can rewrite (\ref{eq:eff_rel}) as:  
\begin{equation} \label{eq:eff_rel2}
\frac{ d\langle c \rangle(t)}{dt} = \mathcal F (\bar{\mu}, \langle c \rangle, \bm b).
\end{equation}

\end{remark}

\section{Exact results for constant diffusion coefficient} \label{sec:exact}
\subsection{Time response to a step load}
Consider an inclusion subjected to a step load, $\bar c(t) = \bar c H(t)$, with $H(t)$ the Heaviside step function. The solution of the linear diffusion problem with constant diffusion coefficient (\ref{eq:Fick2}) can be written under the general form \cite{torquato1991}: 
\begin{equation} \label{eq:local_solution}
\frac{c(\bm x,t)}{\bar c} = 1 - \sum_{n=1}^{\infty} a_n \Psi_n(\bm x)\exp(-\alpha_n t),
\end{equation}
where the eigenfunctions $\Psi_n(\bm x)$ and eigenvalues $\alpha_n$ satisfy  
\begin{equation}
D\bm{\nabla}^2 \Psi + \alpha \Psi = 0, \quad \textnormal{with} \ \Psi=0 \ \textnormal{on} \ \partial \Omega.
\end{equation} 
From (\ref{eq:local_solution}) it is clear that the eigenvalues are inverse relaxation times: $\alpha_n = 1/\tau_n$. The eigenfunctions constitute a complete set of orthogonal functions and are normalised such that:  
\begin{equation} \label{eq:orthonormality}
\langle \Psi_m \Psi_n \rangle = \delta_{mn}.
\end{equation}
The symbol $\langle \cdot \rangle$ was previously introduced in Equation (\ref{eq:vol_av}) and represents a volume average over the inclusion domain. Fulfilment of the initial condition $c(\bm x,0) = 0$ requires that:
\begin{equation}\label{eq:init_cond}
\sum_{n=1}^\infty a_n \Psi_n = 1.
\end{equation}
Using the orthonormality condition (\ref{eq:orthonormality}) of the eigenfunctions together with the condition (\ref{eq:init_cond}), the coefficients $a_n$ are identified:
\begin{equation}\label{eq:an_av}
a_n = \langle \Psi_n \rangle.
\end{equation}
The following property of the coefficients $a_n$ also follows:
\begin{equation}\label{eq:sum_an2}
\sum_{n=1}^{\infty} a^2_n = 1.
\end{equation}

The average concentration in the inclusion volume is directly calculated from the local solution (\ref{eq:local_solution}): 
\begin{equation}\label{eq:average}
\frac{\langle c (t)\rangle}{\bar c} = 1 - \sum_{n=1}^{\infty} A_n \exp(-t/\tau_n), 
\end{equation}
with:
\begin{equation}
A_n \equiv \langle a_n \Psi_n \rangle = a^2_n.  
\end{equation}
The last equality follows directly from the property (\ref{eq:an_av}). From (\ref{eq:sum_an2}), it also follows that
\begin{equation} \label{eq:sum_An}
\sum_{n=1}^{\infty} A_n = 1.
\end{equation}

Introduce the chemical creep function $J(t)$: 
\begin{eqnarray} 
J(t) &\equiv& \frac{\langle c \rangle(t)}{\bar c} \label{eq:creep_fun_def}\\
     &=& 1 - \sum_{n=1}^{\infty} A_n \exp(- t/\tau_n). \label{eq:creep_fun}
\end{eqnarray}
The chemical creep function gives the proportionality ratio between the current average concentration in the inclusion and the constant boundary concentration in a step load. It has the following properties: 
\begin{equation}
\begin{tabular} {rclcl}
$J(0)$ & $=$ & $0$ & \quad &\textnormal{(initial condition)}, \label{eq:init_cond_J}\\
$\lim_{t\to\infty} J(t)$ & $=$ & $1$ &\quad &\textnormal{(chemical equilibrium)}.
\end{tabular}
\end{equation}

\begin{figure}
\begin{center}
\includegraphics[width=0.7\textwidth]{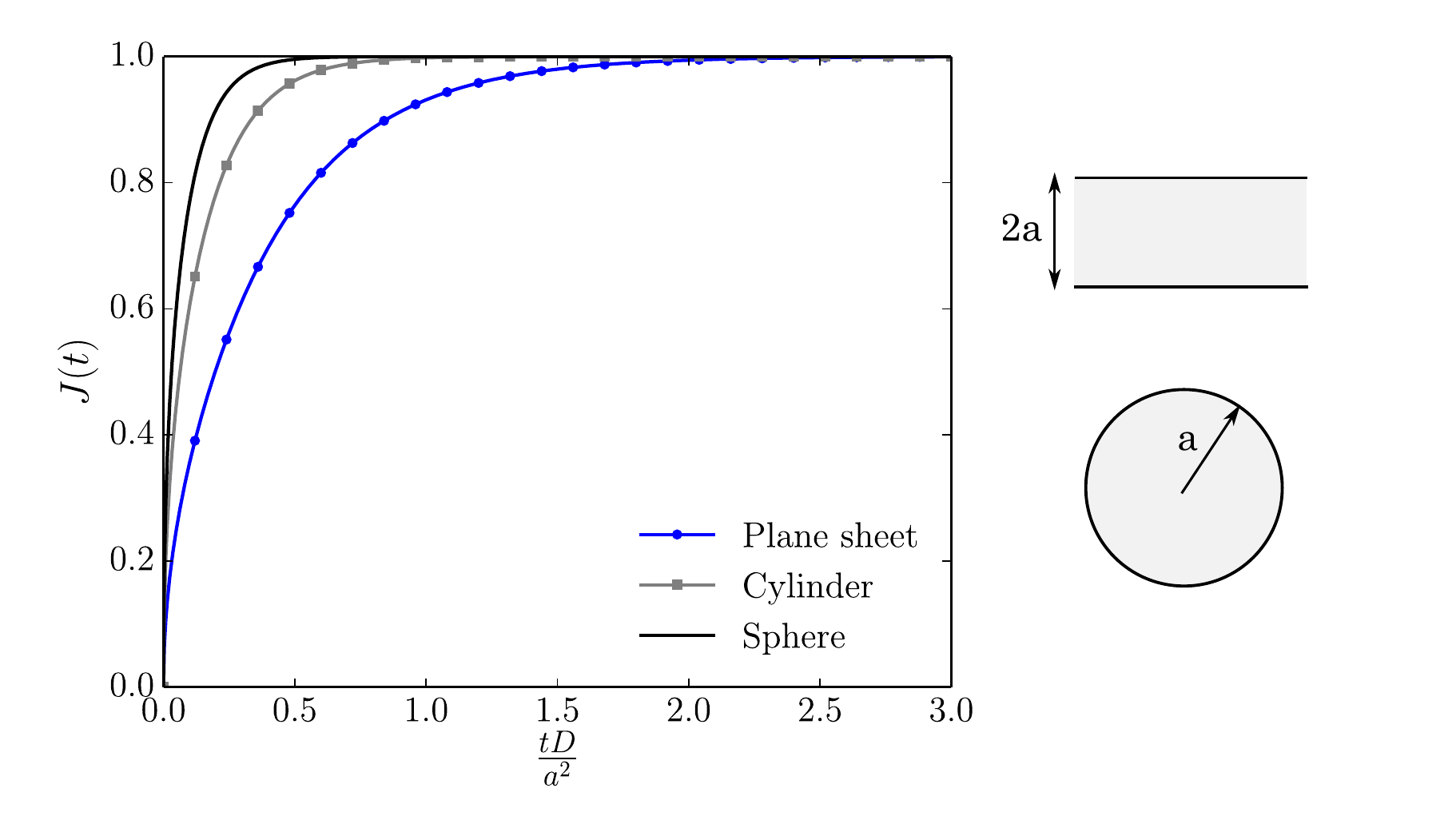}
\caption{Chemical creep function for the linear diffusion problem in a plane sheet, a cylinder and a sphere. The chemical creep functions were generated taking $N=1000$ in the exponential series (\ref{eq:creep_fun}).}
\label{fig:creepfun}
\end{center}
\end{figure}

For simple geometries, solutions to the linear diffusion problem (\ref{eq:Fick2}) can be obtained in closed form using the method of separation of variables \cite{carslaw1959,crank1975}. One-dimensional examples of practical significance are the symmetric diffusion problem through a plane sheet of thickness $2a$, and the radial diffusion problem in a cylinder or a sphere of radius $a$. The coefficients, eigenfunctions and relaxation times for these geometries are given in Table \ref{tab:table1}. In the table, $J_0(z)$ and $J_1(z)$ are Bessel functions of the first kind, and $z_n$ is the n$^{\textnormal{th}}$ root of $J_0$. The corresponding chemical creep functions are illustrated in Figure~\ref{fig:creepfun}.

\begin{table}[h]
\centering
\caption{Solution coefficients for the diffusion problem in a plane sheet of thickness $2a$, a cylinder and a sphere of radius $a$. $r$ is the spatial coordinate. The coefficients $a_n$ are chosen positive and such that the orthonormality condition (\ref{eq:orthonormality}) is satisfied. }
\label{tab:table1}
\begin{tabular}{lccc}
\toprule
& Plane sheet & Cylinder & Sphere \\
\midrule
$a_n$ & $\frac{2\sqrt{2}}{(2n-1)\pi}$ & $\frac{2}{z_n}$ & $\frac{\sqrt{6}}{n\pi}$ \\
$\Psi_n(r)$ & $\sqrt{2} (-1)^{n-1}\cos\left( \frac{(2n-1)\pi r}{2a} \right)$ & $\frac{J_0(\frac{z_n r}{a})}{J_1(z_n)} $ & $\sqrt{\frac{2}{3}}\frac{a}{r}(-1)^{n+1} \sin\left(\frac{n\pi r}{a} \right)$ \\
$\tau_n$ & $\frac{4a^2}{D (2n-1)^2 \pi^2}$ & $\frac{a^2}{D z^2_n}$ & $\frac{a^2}{Dn^2 \pi^2}$\\
$A_n$ & $\frac{8}{(2n-1)^2 \pi^2}$ & $\frac{4}{z^2_n}$ & $\frac{6}{n^2 \pi^2}$ \\
\bottomrule
\end{tabular}
\end{table}

\begin{remark} \label{rmk3}
The function $J(t)$ is called "creep function" in analogy with its mechanical counterpart in  linear viscoelasticity that relates the time-varying strain to the applied stress in a creep experiment. We can extend the analogy by expressing the applied concentration on the boundary in terms of the chemical potential. Introduce $K(\bar{\mu})$ the chemical compliance of the inclusion phase, such that $\bar c = K(\bar{\mu})\bar{\mu}$. Expression (\ref{eq:creep_fun_def}) may be rewritten as:
\begin{equation} \label{eq:creep_fun_mu}
\frac{\langle c \rangle(t)}{\bar{\mu}} = J(t)K(\bar{\mu}).
\end{equation}
There is therefore a correspondence between mechanical and chemical quantities, with stress and chemical potential on the one hand (forces), and strain and (average) concentration on the other hand (fluxes). Note however that the right-hand side of relation (\ref{eq:creep_fun_mu}) is in general function of $\bar{\mu}$, so that the relaxation process is no longer linear when expressed in terms of the $\bar{\mu}$-$\langle c \rangle$ relation. 
\end{remark}

\subsection{Time response to general loading conditions}
The generalisation of the results from the previous section to a general time-dependent boundary condition $\bar c(t)$ relies on Duhamel's theorem, see e.g. \cite{carslaw1959}. According to this principle, the local field in the inclusion subjected to a loading history $\bar c(t)$ is given by
\begin{equation} \label{eq:duhamel}
c(\bm x,t)  = \int_0^t \frac{\partial}{\partial t} F(\bm x,\bar{c}(t'),t-t') dt',
\end{equation}   
where $F(\bm x,\bar{c},t)$ represents the concentration at $\bm x$ at time $t$ in the inclusion  with zero initial concentration and subjected to a step load $ \bar c$ applied in $t=0$. Taking the volume average of (\ref{eq:duhamel}), permuting the time and volume integrals, and introducing the definition of the creep function (\ref{eq:creep_fun_def}), we obtain:
\begin{equation} \label{eq:duhamel_average}
\langle c \rangle(t)  = \int_0^t  \frac{\partial}{\partial t} J(t-t') \bar c(t') dt'.
\end{equation}   
Using the expression of the creep function (\ref{eq:creep_fun}) and integrating by parts, on finally obtains: 
\begin{equation} \label{eq:integral_representation}
\langle c \rangle(t) = \bar c(0)J(t) + \int_0^t J(t-t') \frac{d \bar c}{dt'} dt'.   
\end{equation}   
The first term on the right-hand side of Equation (\ref{eq:integral_representation}) represents the average concentration response to a step load applied in $t=0$, and the second term represents the response due to the subsequent time evolution of the boundary condition. The representation (\ref{eq:integral_representation}) is similar to the Boltzmann superposition principle in linear viscoelasticity: the average concentration in the body in response to a time-dependent concentration boundary condition is given by the superposition of the responses to individual incremental step loads $\Delta \bar c$. When $\bar c(0)=0$, the integral representation (\ref{eq:integral_representation}) reduces to:
\begin{equation} \label{eq:integral_representation2}
\langle c \rangle(t) =  \int_0^t J(t-t') \frac{d\bar c}{dt'} dt'.   
\end{equation}      

The effective, steady-state response of an inclusion subjected to harmonic loading is discussed in the appendix, where we also introduce the concept of chemical storage and loss moduli in analogy with the corresponding quantities in viscoelasticity.

\section{Estimates of the chemical creep function for arbitrary geometries } \label{sec:estimates}
In the last section we showed that the effective response of an inclusion subjected to a uniform, time-varying Dirichlet boundary condition was fully characterised by the chemical creep function (\ref{eq:creep_fun}) expressed as infinite series of decaying exponentials with coefficients $A_n$ and relaxation times $\tau_n$. Except for the case of simple geometries, it is in general not possible to obtain the coefficients and relaxation times in closed form, and a numerical resolution of the boundary value problem is required.

In this section we propose several methods to generate estimates of the creep function written as Prony series having a \textit{finite} number $N$ of relaxation times. Based on the exact representation (\ref{eq:creep_fun}), we are looking for estimates of the form:        
\begin{eqnarray} 
\tilde J (t)  &=& 1 - \sum_{n=1}^{N} \tilde A_n \exp(- t/\tilde{\tau}_n)  \label{eq:creep_fun_approx1}\\
              &=& \sum_{n=1}^N \tilde A_n \left( 1 - \exp(-t/\tilde{\tau}_n) \right) + \tilde A_{N+1}. \label{eq:creep_fun_approx2}
\end{eqnarray}
Here, $\tilde{\tau}_n$ are the relaxation times and $\tilde A_n$ the coefficients of the approximate creep function. In (\ref{eq:creep_fun_approx2}), the coefficient $\tilde A_{N+1}$ is defined such that  $\sum_{n=1}^{N+1} \tilde A_n = 1$, that is, $\tilde A_{N+1} = 1-\sum_{n=1}^N \tilde A_n$. The representation (\ref{eq:creep_fun_approx2}) is akin to a Prony series representation of creep and relaxation functions in viscoelasticity. The estimate (\ref{eq:creep_fun_approx1})-(\ref{eq:creep_fun_approx2}) has the following properties: 
\begin{equation}
\begin{tabular} {rclcl}
$\tilde J(0)$ & $=$ & $ \tilde A_{N+1}$ & \quad &\textnormal{(initial condition)},\\
$\lim_{t\to\infty} \tilde J(t)$ & $=$ & $1$ &\quad &\textnormal{(chemical equilibrium)}.
\end{tabular}
\end{equation}
In general the estimate (\ref{eq:creep_fun_approx1})-(\ref{eq:creep_fun_approx2}) predicts a non-zero instantaneous average concentration in response to a step load, unless the coefficients $\tilde A_n$ are chosen such that $\sum_{n=1}^N \tilde A_n = 1$. On the other hand, it correctly predicts the long-time equilibrium response by construction. 

The estimate (\ref{eq:creep_fun_approx1})-(\ref{eq:creep_fun_approx2}) of the creep function can be used in the integral expression (\ref{eq:integral_representation}) in order to predict the inclusion response under an arbitrary loading history $\bar c(t)$. In the next section, we propose an internal variable representation that is much more amenable to numerical time integration than the integral representation (\ref{eq:integral_representation}). Particular methods to identify the coefficients $\tilde A_n$ and $\tilde{\tau}_n$ are presented in Sections \ref{sec:colloc}, \ref{sec:opti} and \ref{sec:modal}. 

\subsection{Equivalent representation of the effective inclusion response based on internal variables} \label{sec:diff_rep}
Since the approximate creep function (\ref{eq:creep_fun_approx1})-(\ref{eq:creep_fun_approx2}) has a finite number of relaxation times, it is possible to describe the effective inclusion response using a finite number of internal variables \cite{mandel1966,ricaud2009}. Introduce the approximate creep function (\ref{eq:creep_fun_approx2}) into (\ref{eq:integral_representation}). By inspection, we can decompose the average concentration into $N+1$ internal variables: 
\begin{equation} \label{eq:modes}
\langle c \rangle  = \sum_{n=1}^{N} b_n + b_{N+1}.
\end{equation}
The variables $b_n$ ($n=1,N$) evolve according to: 
\begin{equation}\label{eq:bn}
b_n(t) = \bar c(0)\tilde A_n \exp(-t/\tilde{\tau}_n) + \int_0^t \tilde A_n (1 - \exp(-(t-t')/\tilde{\tau}_n) \frac{d\bar c}{dt'} dt',\quad n=1,N 
\end{equation}
The expression (\ref{eq:bn}) is solution to an ODE for $b_n$: 
\begin{equation} \label{eq:ODE_bn}
\tilde{\tau}_n \dot{b}_n + b_n = \tilde A_n \bar c. 
\end{equation}
with the initial condition $b_n(0) = 0$. On the other hand, the variable $b_{N+1}(t)$ evolves according to:
\begin{equation} \label{eq:bnp1}
b_{N+1} = \tilde A_{N+1} \bar c .
\end{equation}
The ODEs (\ref{eq:ODE_bn}) and (\ref{eq:bnp1}), together with (\ref{eq:modes}) completely characterise the effective inclusion response. The sought-after expression (\ref{eq:eff_rel}) is thus formally identified as:
\begin{equation}\label{eq:eff_model1}
\frac{d\langle c \rangle}{dt} = \sum_{n=1}^N \frac{\tilde A_n \bar c - b_n}{\tilde{\tau}_n} + \tilde A_{N+1} \frac{d\bar c}{dt}.
\end{equation}
In practice, the effective inclusion response under an arbitrary loading history $\bar c(t)$ is readily calculated by integrating  the linear ODE's (\ref{eq:ODE_bn}) and (\ref{eq:bnp1}) in time using a fully implicit Euler scheme.

\begin{figure}
\centering
\includegraphics[width=0.6\textwidth]{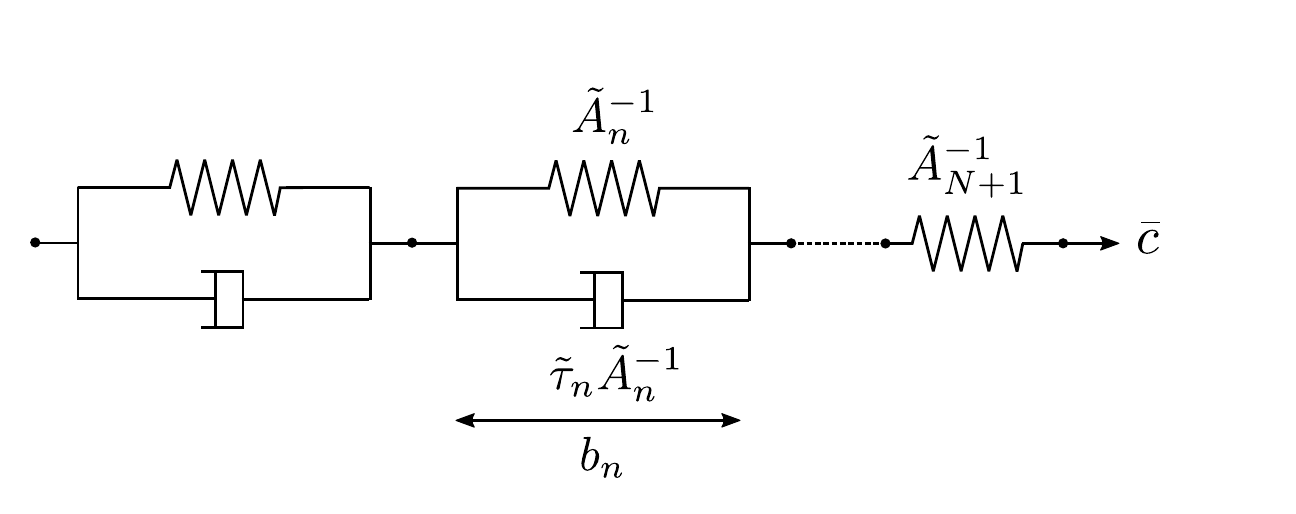}
\caption{Models for the average concentration response of an inclusion subjected to a uniform, time-varying concentration boundary condition can be represented by a series of Kelvin-Voigt units and an additional spring. The average concentration $\langle c \rangle$ is represented by the total elongation of the assembly, obtained by summing the elongations $b_n$ in each spring/dashpot unit. }
\label{fig:KV}
\end{figure}

The effective constitutive model (\ref{eq:ODE_bn})-(\ref{eq:bnp1}) admits a simple representation based on springs and dashpots, as commonly used in phenomenological models of viscoelasticity. Here, the model consists of an assembly of $N$ Kelvin-Voigt units in series with one spring, where each concentration-like internal variable $b_n$ ($n=1,N$)  is the elongation of the n$^{\textnormal{th}}$ unit with spring stiffness $\tilde A^{-1}_{n}$ and dashpot viscosity $\tilde{\tau}_n \tilde A^{-1}_{n}$, as represented in Figure~\ref{fig:KV}. The additional  spring with stiffness $\tilde A^{-1}_{N+1}$ describes the instantaneous response under a step load.   

\begin{remark}
We can express the effective inclusion response in a form similar to (\ref{eq:eff_rel2}) for a given history of chemical potential $\bar{\mu}(t)$ applied on its boundary. Introducing the chemical compliance function, the model (\ref{eq:eff_model1}) becomes 
\begin{equation}\label{eq:eff_model2}
\frac{d\langle c \rangle}{dt} = \sum_{n=1}^N \frac{\tilde A_n K(\bar{\mu}) \bar{\mu} - b_n}{\tilde{\tau}_n} + \tilde A_{N+1} \frac{d(K(\bar{\mu})\bar{\mu})}{dt}.
\end{equation}
Inclusion problems expressed in terms of applied chemical potential on the inclusion surface require the inversion of the (generally non-linear) chemical potential-concentration relation at each time step in order to calculate the right-hand side of (\ref{eq:ODE_bn})-(\ref{eq:bnp1}). However, this does not affect the time integration algorithm to update the internal variables. 
\end{remark}

\subsection{Collocation method} \label{sec:colloc}
A first, simple method to estimate the amplitudes and relaxation times in (\ref{eq:creep_fun_approx1}) is the collocation method initially proposed by Schapery \cite{schapery1961} in the context of linear viscoelasticity for fitting Prony series to experimental creep or relaxation curves. The collocation method is also used in the context of homogenisation of linear viscoelastic composites to perform numerical inversion of the Laplace-Carson (LC) transform \cite{masson1999,pierard2006}. Here we use the collocation method directly in the time domain. The method requires a preliminary determination of the "exact" creep function for the inclusion subjected to a step load $\bar{c}$ applied at $t=0$, typically using the FE method or an analytical approach, if available.   

Following Schapery \cite{schapery1961}, we choose the $N$ relaxation times $\tilde{\tau}_n$ to be equispaced on a logarithmic scale between some arbitrarily chosen minimum and maximum times $\tau_{min}$ and $\tau_{max}$. Let $a^2/D$ be the time scale for diffusion, with $a$ the characteristic size of the inclusion. We may assume that diffusion is negligible when the observation time is much smaller than the characteristic diffusion time, and completely relaxed when the observation time is larger than the characteristic diffusion time. These considerations allow us to narrow the range of relevant relaxation times in the collocation method. The amplitudes associated with each relaxation time are then calculated in such a way that the approximate creep function coincides with the reference one at the pre-defined collocation times: $J(\tilde{\tau}_k) = \tilde{J}(\tilde{\tau}_k)$. For estimates of the form (\ref{eq:creep_fun_approx1}), this conditions rewrites as:
\begin{equation}
J(\tilde{\tau}_k) = 1 - \sum_{n=1}^N \tilde A_n \exp(-\tilde{\tau}_k/\tilde{\tau}_n), \quad k=1,N.
\end{equation}
The method provides a system of $N$ linear equations for the $N$ unknown amplitudes $\tilde A_n$, while $\tilde A_{N+1}$ is calculated from its definition: $\tilde A_{N+1}=1-\sum_{n=1}^N \tilde A_n$. The advantage of the method is its simplicity and ease of implementation. A major drawback, however, is that the calculated amplitudes can be negative, which leads to spurious modes in the series expansion of the creep function.  In addition, the method does not satisfy the condition $\sum_{n=1}^N \tilde A_n =1$, and therefore predicts an  instantaneous average concentration in response to a step load. Finally, the relaxation times need to be determined a priori and are therefore not optimised.   

\subsection{Constrained optimisation method}\label{sec:opti}
The shortcomings of the collocation method can be addressed by relying on an optimisation approach under constraints to fit the coefficients in the Prony series (\ref{eq:creep_fun_approx1}). Here we took  inspiration from the optimised collocation method proposed by Rekik and Brenner in the context of viscoelasticity \cite{rekik2011}. First define the error between the reference ("exact") and approximate creep functions as:  
\begin{equation}\label{eq:err_fun}
e(\tilde A_n, \tilde{\tau}_n; N_p) = \sum_{i=1}^{N_p} \left( \tilde J(t_i) - J(t_i) \right)^2,
\end{equation}
where the summation is taken over $N_p$ pre-selected data points. The error is a function of the coefficients in the series expansion, as well as the number of chosen data point as a parameter. The optimisation problem consists in identifying the set of coefficients $\left\{\tilde A_n, \tilde{\tau}_n\right\}$ that minimises the error function (\ref{eq:err_fun}) under the following constraints:
\begin{itemize}
\item All the coefficients $\tilde A_n$ should be non-negative: $\tilde A_n \geq 0$, $n=1,N$.
\item The creep function should predict a zero instantaneous concentration: $\sum_{n=1}^N \tilde A_n = 1$;
\item The relaxation times $\tilde{\tau}_n$ are positive and should lie within some interval $[\tau_{min},\tau_{max}]$.
\end{itemize}
The least-square minimisation under constraint was performed using the Sequential Least Square Programming of the Python optimisation module pyOpt \cite{perez2012}, which relies the Han-Powell quasi-Newton method. As illustrated in the next section, the method can provide a very accurate estimate of the creep function for a very small number of relaxation times. The drawback, however, is that the method is very sensitive to the choice of initial values of the unknown coefficients $\tilde A_n$ and $\tilde{\tau}_n$.  

\subsection{Method based on FE modal analysis}\label{sec:modal}
A third method relies on a modal analysis of the FE solution of the inclusion subjected to a uniform and constant Dirichlet boundary condition $\bar{c}$ applied at $t=0$, see \cite{curnier1994} for details. The concentration field is discretized as:
\begin{equation} \label{FEdiscr}
\frac{c(x,t)}{\bar{c}} = \sum_{I=1}^{N_n} N_I(x)u_I(t),
\end{equation} 
with $N_n$ the number of nodes in the FE discretisation, $N_I(x)$ the shape function associated with node $I$, and $u_I(t)$ the nodal value at node $I$. For conciseness, we write the nodal values in a vector $\bm u$. After discretising the weak form of equation (\ref{eq:Fick2}) using the standard Galerkin approach, the vector of nodal values $\bm u$ is found to obey the following system:  
\begin{equation} \label{eq:FE_sys}
\bm C \cdot \dot{\bm u} + \bm D \cdot \bm u = \bm 0 ,
\end{equation}   
with $\bm C$ the chemical capacity matrix and $\bm D$ the conductivity matrix:
\begin{eqnarray}
C_{IJ} &=& \int_{\Omega} N_I(\bm x) N_J(\bm x) \ dV, \label{capacity} \\
D_{IJ} &=& D\int_{\Omega} \bm{\nabla} N_I(\bm x) \cdot \bm{\nabla} N_J(\bm x) \ dV. \label{diffusivity}
\end{eqnarray}
The vector $\bm u(t)$ solution of (\ref{eq:FE_sys}) is of the form:
\begin{equation} \label{nodal_solution}
\bm u(t) = \bm 1 - \sum_{n=1}^{N_n} a_n \bm y_n \exp(-\alpha_n t),
\end{equation}
where $\bm 1$ is a vector of size $N_n$ with all elements equal to one, such that $\bm D \cdot \bm 1 = \bm 0$,
and where $\bm y_n$ and $\alpha_n$ are the eigenvectors and eigenvalues, solution of:
\begin{equation} \label{eq:eigenproblem_FE}
[\bm C^{-1}\bm D] \cdot \bm y = \alpha \bm y .
\end{equation} 
The eigenvectors are such that:
\begin{eqnarray}
\bm y^T_m \cdot \bm C \cdot \bm y_n &=& V(\Omega)\delta_{mn} \label{eq:orthogonality1} \\ 
\bm y^T_m \cdot \bm D \cdot \bm y_n &=& V(\Omega)\alpha_n \delta_{mn} \label{eq:orthogonality2}
\end{eqnarray}
The coefficients $a_n$ are identified from the initial condition $\sum_{n=1}^{N_n} a_n \bm y_n = \bm 1$. Pre-multiplying each side of the latter equation by $\bm y^T_m \cdot \bm C$ and using the orthogonality condition (\ref{eq:orthogonality1}):
\begin{equation} \label{eq:an_FEM}
a_n = \frac{1}{V(\Omega)}\bm y^T_n  \cdot \bm C \cdot \bm 1 .
\end{equation} 
The factor $V(\Omega)$ in equations (\ref{eq:orthogonality1}), (\ref{eq:orthogonality2}) and (\ref{eq:an_FEM}) was introduced to be consistent with (\ref{eq:orthonormality}). 

In practice, the method requires the solution of the eigenvalue problem (\ref{eq:eigenproblem_FE}) for the modes $\bm y_n$ and eigenvalues $\alpha_n$. The coefficients $a_n$ are then calculated from (\ref{eq:an_FEM}). The coefficients in the estimate (\ref{eq:creep_fun_approx1}) are finally given by $\tilde A_n = a_n^2$, $\tilde{\tau}_n=1/\alpha_n$. The number of modes $N$ in the estimate can be chosen such that  $1 \leq N \leq N_n$. In the following examples, we sorted the modes by descending order based on the amplitude $\tilde A_n$.   

The application of the FE modal analysis approach is straightforward and does not require any optimisation or initial guess. It does require, however, to have access to the capacity and diffusivity matrices of the FE system, which may not be straightforward if a commercial FE software is used.

\section{Results} \label{sec:results}

\subsection{Verification of the estimates in a one-dimensional case}
We verify the accuracy of the different estimates proposed in Section \ref{sec:estimates} in the example of radial diffusion in a cylinder of radius $a$, for which an exact analytical expression of the chemical creep function exists. The reference analytical solution was previously shown in Figure~\ref{fig:creepfun} and is obtained by taking the first 1000 modes in the Prony series (\ref{eq:creep_fun}) with coefficients reported in Table \ref{tab:table1}.

Estimates based on the simple collocation method (Method 1) were obtained setting  $\tau_{min}=10^{-6}a^2/D$ and $\tau_{max} = a^2/D$. For the constrained optimisation method (Method 2), we have chosen $N_p=10$ data points that are equispaced on a linear time scale between 0 and $a^2/D$. The minimum relaxation time $\tau_{min}$ was set to $10^{-6}a^2/D$ and the maximum relaxation time $\tau_{max}$ to $10 a^2/D$. We chose the exact, analytical modes and relaxation times for each mode as first guess in the optimisation procedure. Finally, the capacity and conductivity matrices for the FE modal analysis (Method 3) were calculated using linear shape functions and 101 equispaced nodes along the radius. Spatial integration in Equations (\ref{capacity})-(\ref{diffusivity}) was carried out exactly, taking into account axisymmetry. 

\begin{figure}
\begin{center}
\begin{subfigure}{0.45\textwidth}
\includegraphics[width=\textwidth]{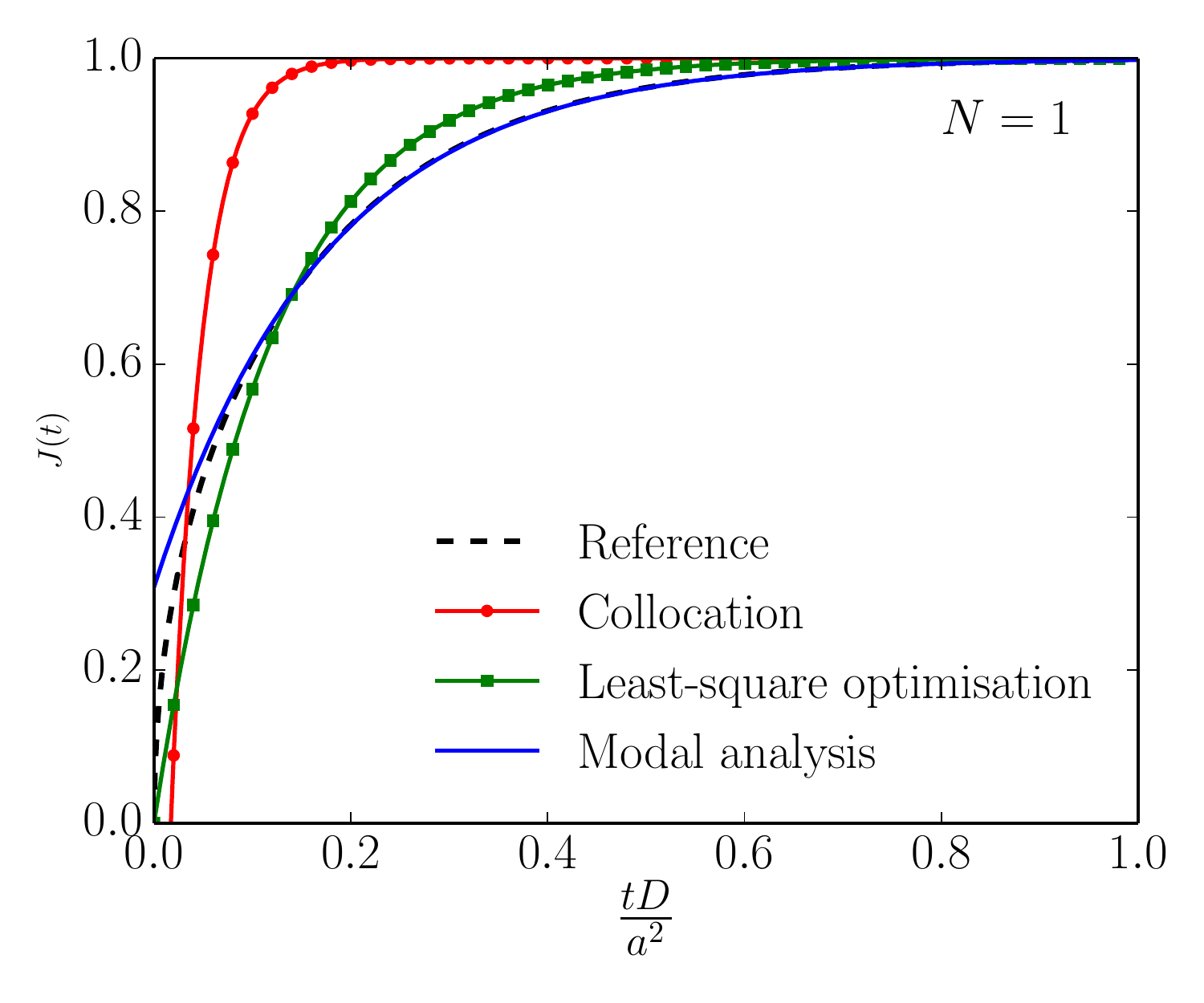}
\caption{}
\end{subfigure}
\begin{subfigure}{0.45\textwidth}
\includegraphics[width=\textwidth]{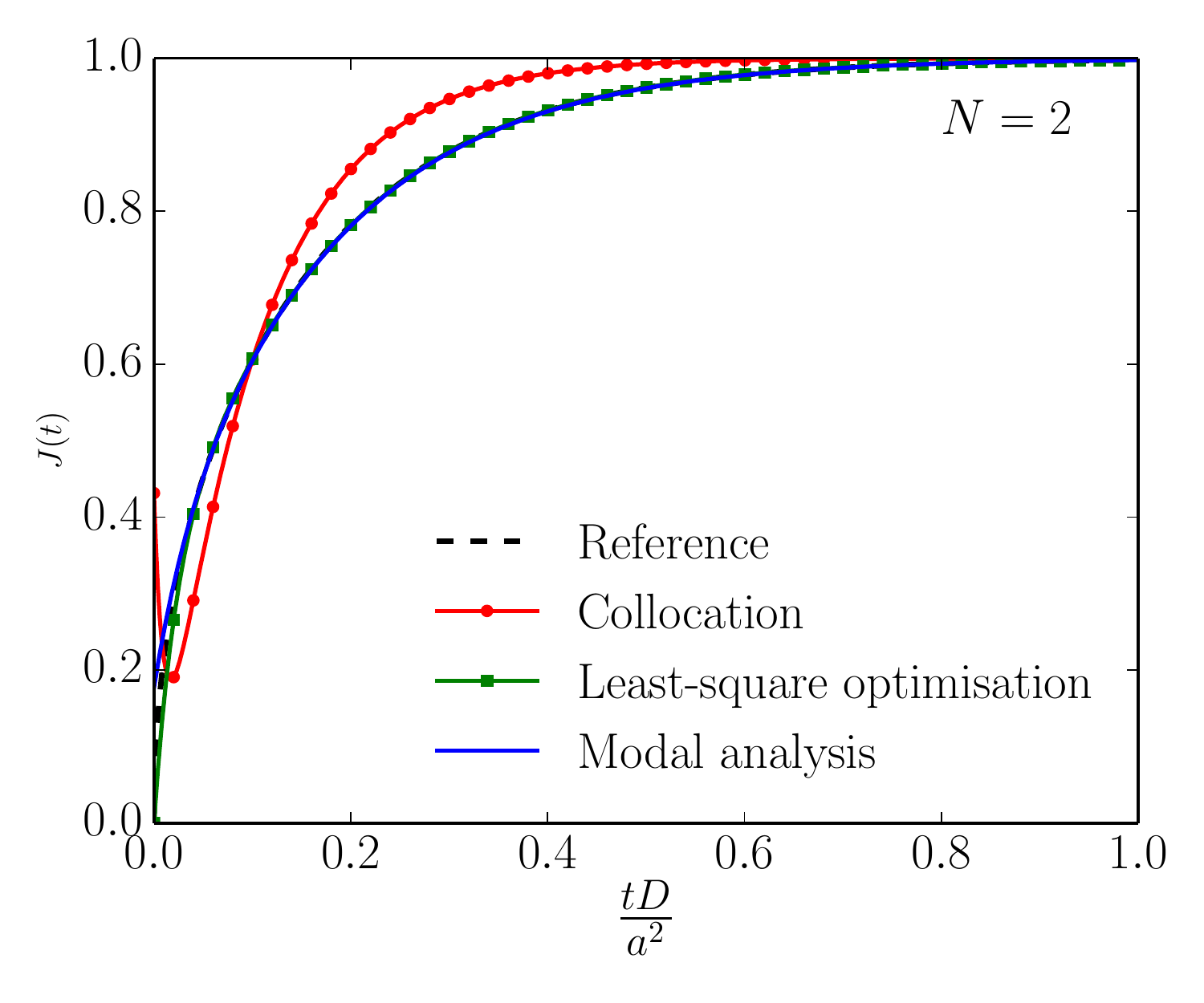}
\caption{}
\end{subfigure}
\begin{subfigure}{0.45\textwidth}
\includegraphics[width=\textwidth]{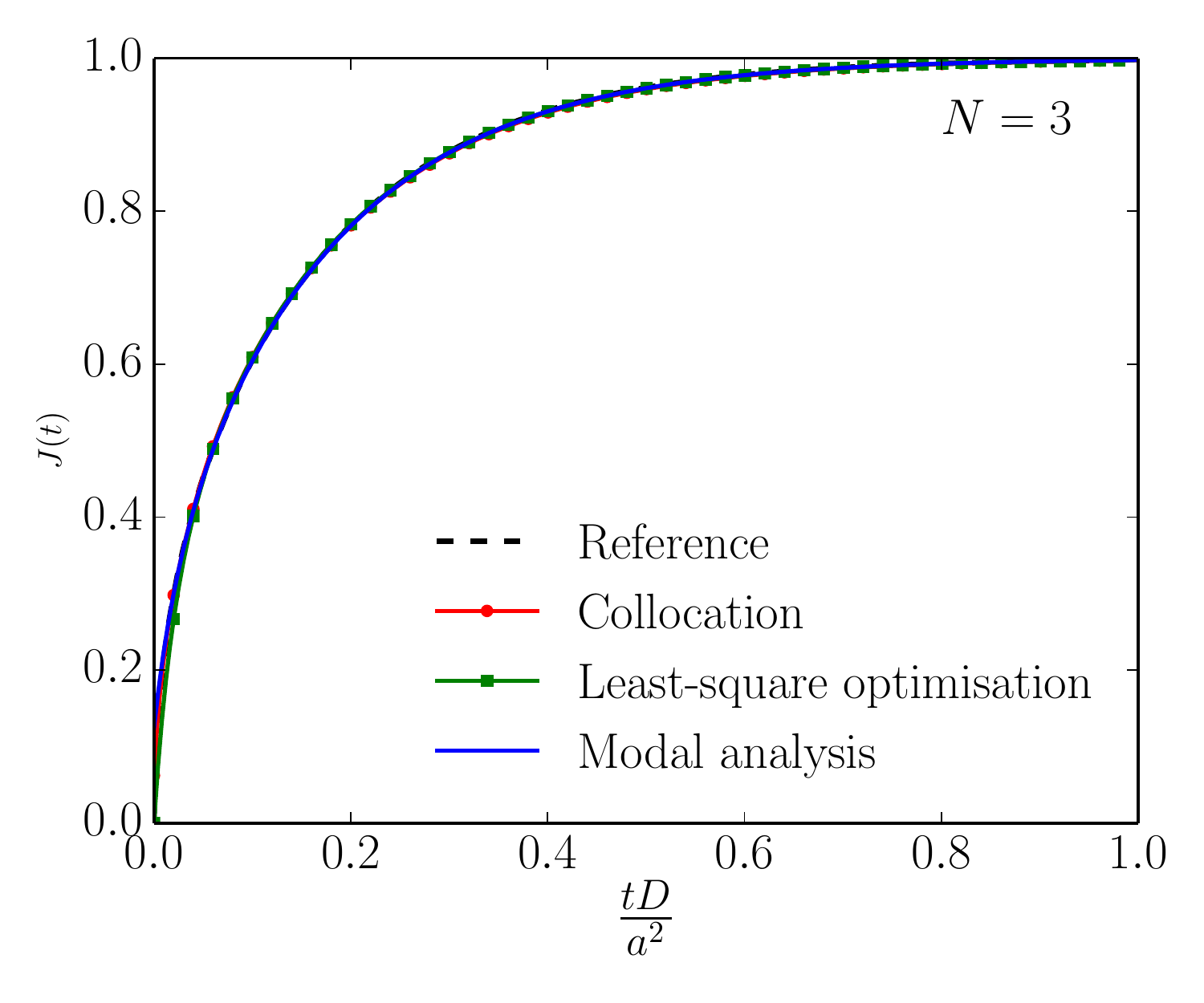}
\caption{}
\end{subfigure}
\begin{subfigure}{0.45\textwidth}
\includegraphics[width=\textwidth]{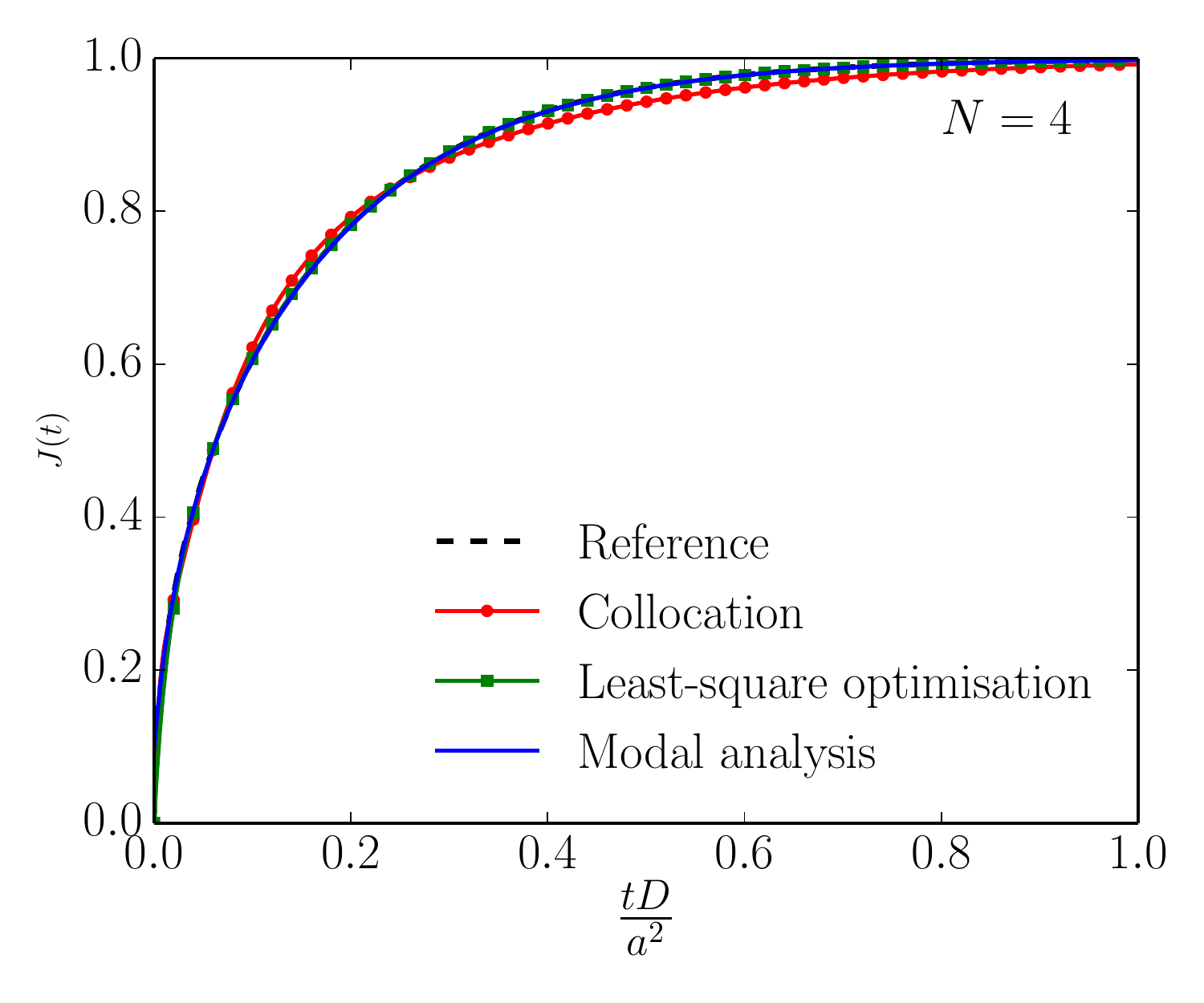}
\caption{}
\end{subfigure}
\caption{Chemical creep function for the radial diffusion problem in a cylinder as estimated by three different methods: a collocation method, a constrained optimisation method, and a FE-based modal analysis approach, and for various numbers of modes $N$. The reference curve is obtained using 1000 terms in the series expansion (\ref{eq:creep_fun}), with the analytical coefficients reported in Table \ref{tab:table1}.}
\label{fig_cylinder}
\end{center}
\end{figure}

The chemical creep function as estimated with the different methods and increasing number of modes $N$ is illustrated in Figure~\ref{fig_cylinder}. By construction, all the estimates predict the correct long-time behaviour. In contrast, only the constrained optimisation approach also satisfies the short-time requirement (\ref{eq:init_cond_J}). The accuracy of all these methods increases as the number of modes $N$ increases, as expected. 

The collocation method is the simplest and cheapest method among the three. However, it may yield negative coefficients $\tilde{A}_n$, leading to non-monotonic curves for some value of $N$, as for example for $N=2$, Figure \ref{fig_cylinder}b. The constrained optimisation method gives a very good estimate, including for very small number of modes. However, the quality of the optimisation is highly dependent on the initial guess for the coefficients and relaxation times. In this particular example we could rely on the exact analytical solutions $A_n$ and $\tau_n$ as initial guesses for $\tilde A_n$ and $\tilde{\tau}_n$, and the optimisation method thus improves the estimate based on a simple truncation of the exact series expansion (see also below). In contrast, arbitrary initial guesses may lead to very poor estimates (not shown). This is a major drawback in the general case where the Prony series expansion of the reference curve is not known analytically. 

Finally, the FE modal analysis approach provides a good compromise between the former two approaches. The coefficients and relaxation times obtained from FE modal analysis are reported in Table \ref{table_modal} and compared to their analytical counterparts, up to $N=5$. The coefficients and relaxation times are very close, and the modal analysis approach gives practically the same estimate as the simple truncation of the exact series expansion (\ref{eq:creep_fun}) (not represented in Figure \ref{fig_cylinder}) in this example.

\begin{table}[h]
 \centering
 \caption{First five coefficients $A_n$ and relaxation times $\tau_n$ for the radial diffusion problem in a cylinder, as obtained from the exact solution (Table \ref{tab:table1}) and the FE-based modal analysis with 100 elements and linear shape functions.}
 \label{table_modal}
\begin{tabular}{|l|c|c|c|c|}
 \hline
$N$ & \multicolumn{2}{|c|}{$A_n$}     &   \multicolumn{2}{|c|}{$\tau_n$}         \\ \hline
    & Modal & Analytical & Modal & Analytical \\ \hline 
 
    1    &  \num{6.91409e-01} & \num{6.91660e-01} & \num{1.73916e-01} & \num{1.72915e-01}        \\    
    2    &  \num{1.31316e-01} & \num{1.31271e-01} & \num{3.30460e-02} & \num{3.28178e-02}	  	   \\
    3    &  \num{5.34518e-02} & \num{5.34138e-02} & \num{1.34509e-02} & \num{1.33535e-02}		   \\
    4    &  \num{2.87965e-02} & \num{2.87686e-02} & \num{7.24427e-03} & \num{7.19216e-03}        \\
    5    &  \num{1.79634e-02} & \num{1.79427e-02} & \num{4.51664e-03} & \num{4.48567e-03}         \\ \hline
\end{tabular}
\end{table}

\subsection{Application to arbitrary 2D geometries and general loading conditions}
\begin{figure}
\begin{center}
\begin{subfigure}{0.6\textwidth}
\includegraphics[width=\textwidth]{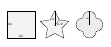}
\caption{}
\end{subfigure}
\begin{subfigure}{0.45\textwidth}
\includegraphics[width=\textwidth]{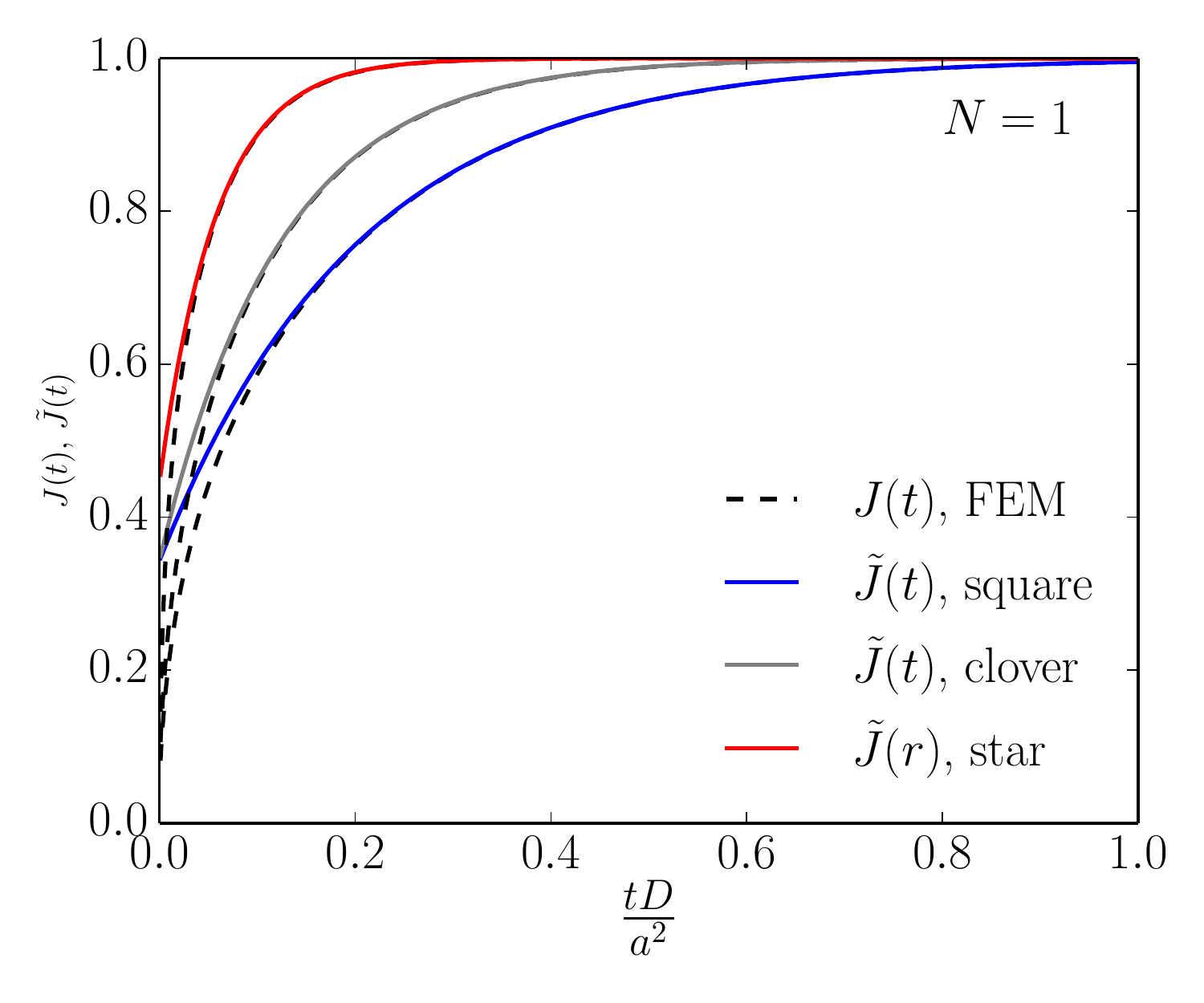}
\caption{}
\end{subfigure}
\begin{subfigure}{0.45\textwidth}
\includegraphics[width=\textwidth]{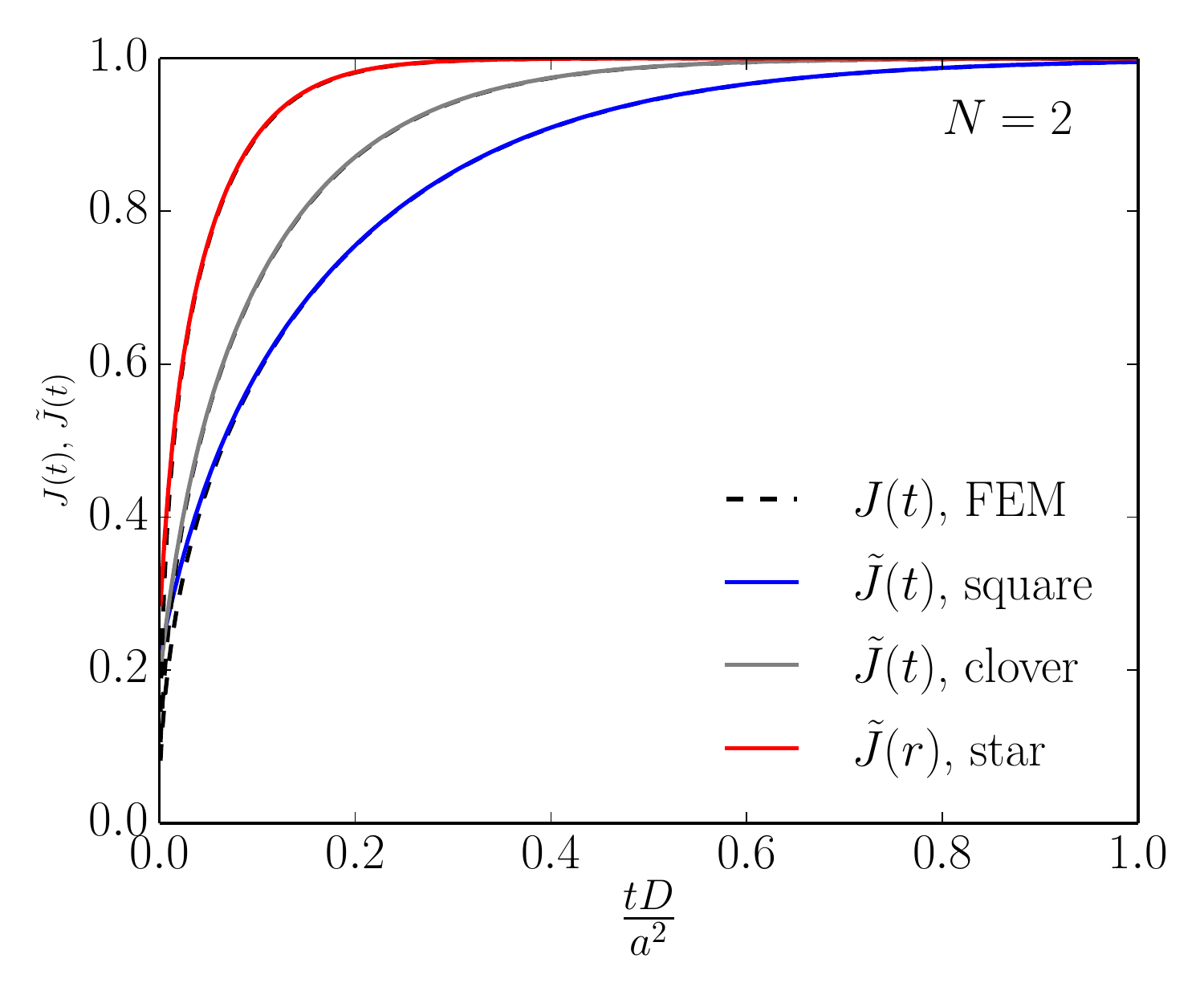}
\caption{}
\end{subfigure}
\begin{subfigure}{0.45\textwidth}
\includegraphics[width=\textwidth]{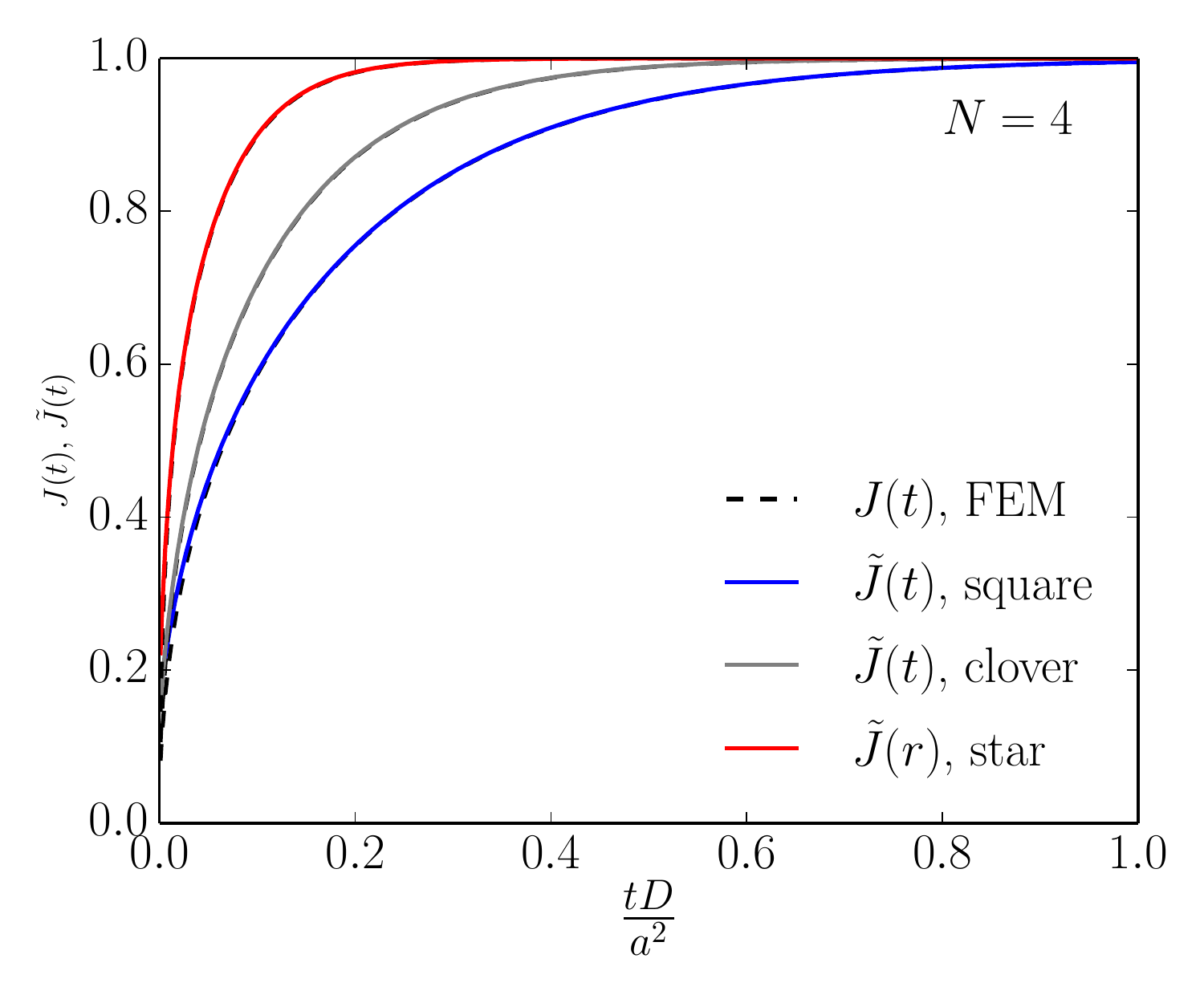}
\caption{}
\end{subfigure}
\begin{subfigure}{0.45\textwidth}
\includegraphics[width=\textwidth]{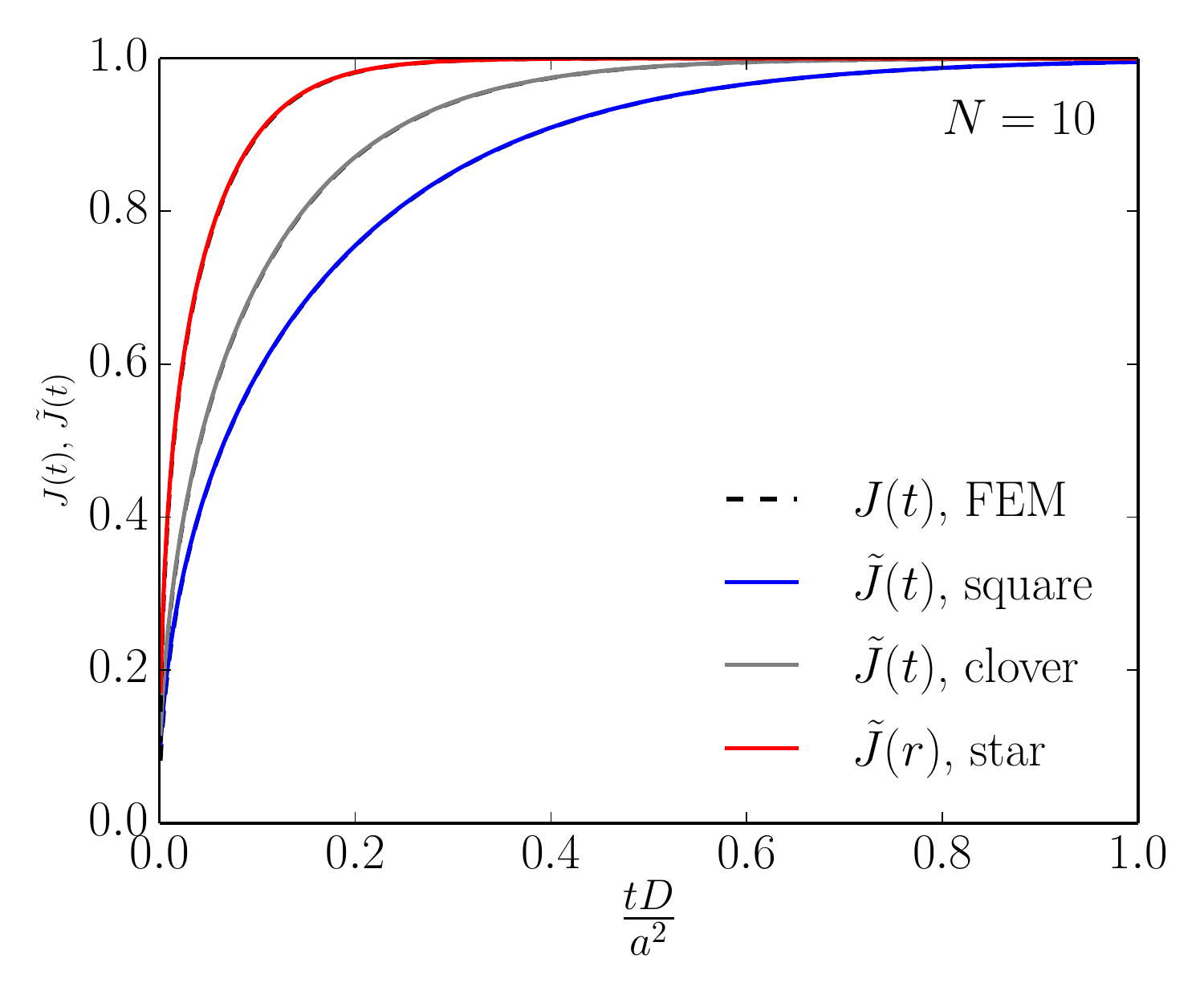}
\caption{}
\end{subfigure}
\caption{(a) The considered 2D geometries with square, star and clover shapes, and (b)-(e) the corresponding creep functions.  Reference creep functions $J(t)$ were calculated from the full-field FE solution, while the estimates $\tilde J(t)$ were obtained using FE modal analysis. }
\label{fig:2Dgeom}
\end{center}
\end{figure}

We now address the diffusion problem in general 2D geometries, for which an analytical expression of the coefficients in the series expansion (\ref{eq:creep_fun}) is generally not available. The considered geometries are represented in Figure \ref{fig:2Dgeom}a, which also shows the characteristic length $a$. In these examples, the reference creep function was obtained by solving the diffusion problem under a step load using the FE method. Meshing was done with the software Gmsh \cite{gmsh}, and we used an in-house FE code. Second-order triangular elements were used, and the number of elements was about 500. The reference creep function was directly calculated from the ratio $\langle c \rangle(t)/\bar c$. 

FE modal analysis was used to estimate the creep function with a limited number of modes. The same FE mesh and shape functions were used for both the reference and modal analysis solutions, and the integrals (\ref{capacity}) and (\ref{diffusivity}) were evaluated numerically. Figures \ref{fig:2Dgeom}b-e compare the reference creep functions to their estimates based on FE modal analysis, for the considered geometries. As expected, the error is mainly on the short-time response, and decreases as the number of modes increases.

Next we use the coefficients $\tilde A_n$ and relaxation times $\tilde{\tau}_n$ obtained from FE modal analysis on these geometries to \textit{predict} the average concentration response under general time evolution of the boundary condition. To this end, we use the internal variable representation described in Section \ref{sec:diff_rep}, and update the internal variables $b_n$ using  (\ref{eq:ODE_bn}) and (\ref{eq:bnp1}). The predictions of this semi-analytical approach are compared to reference predictions obtained from full-field FE simulations of the inclusion response subject to time-varying loading conditions. In these examples, the calculations based on the semi-analytical estimates of the creep function are orders of magnitude cheaper than the full-field FE calculations.  

\begin{figure}
\begin{center}
\begin{subfigure}{0.45\textwidth}
\includegraphics[width=\textwidth]{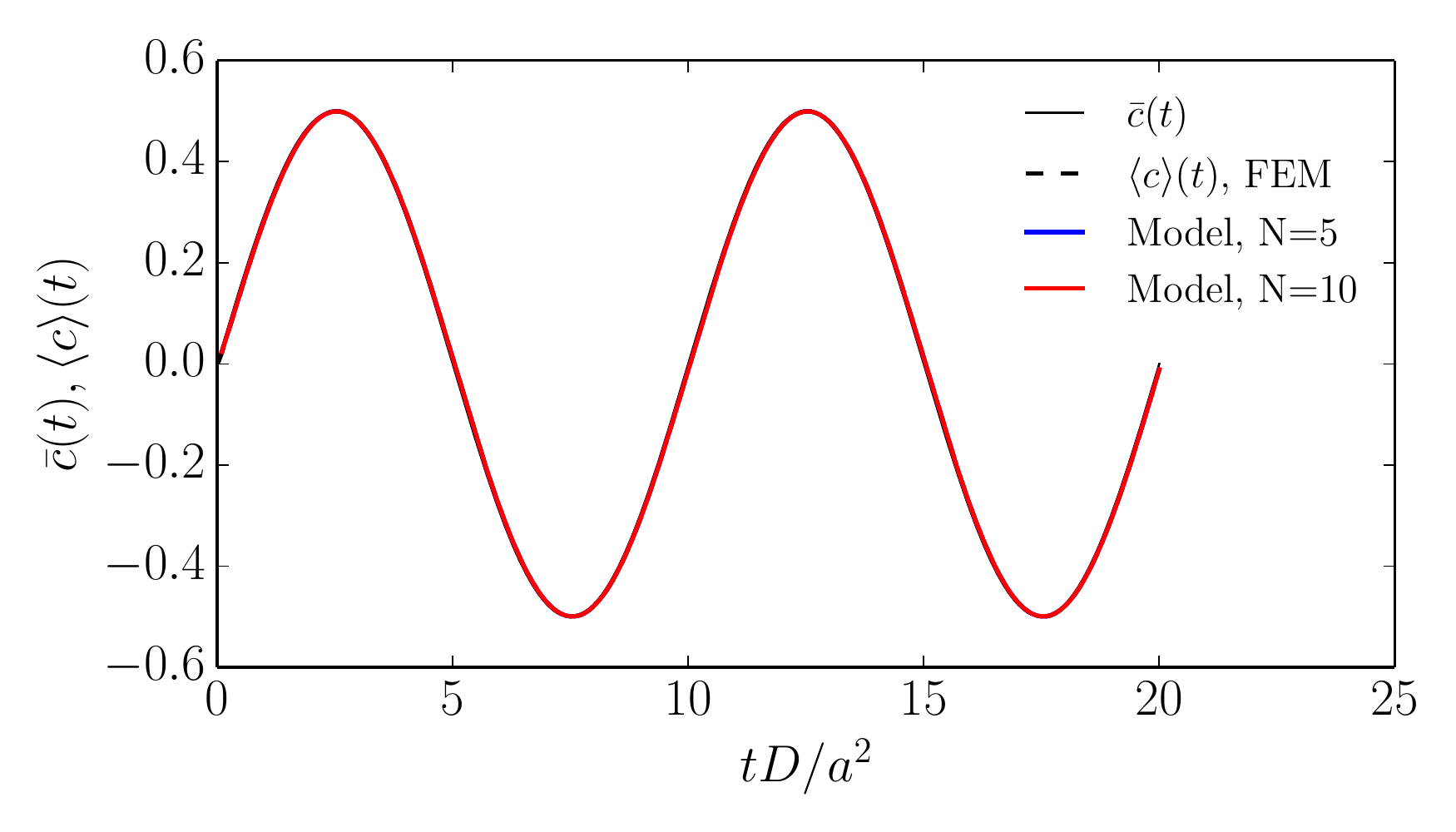}
\caption{}
\end{subfigure}
\begin{subfigure}{0.45\textwidth}
\includegraphics[width=\textwidth]{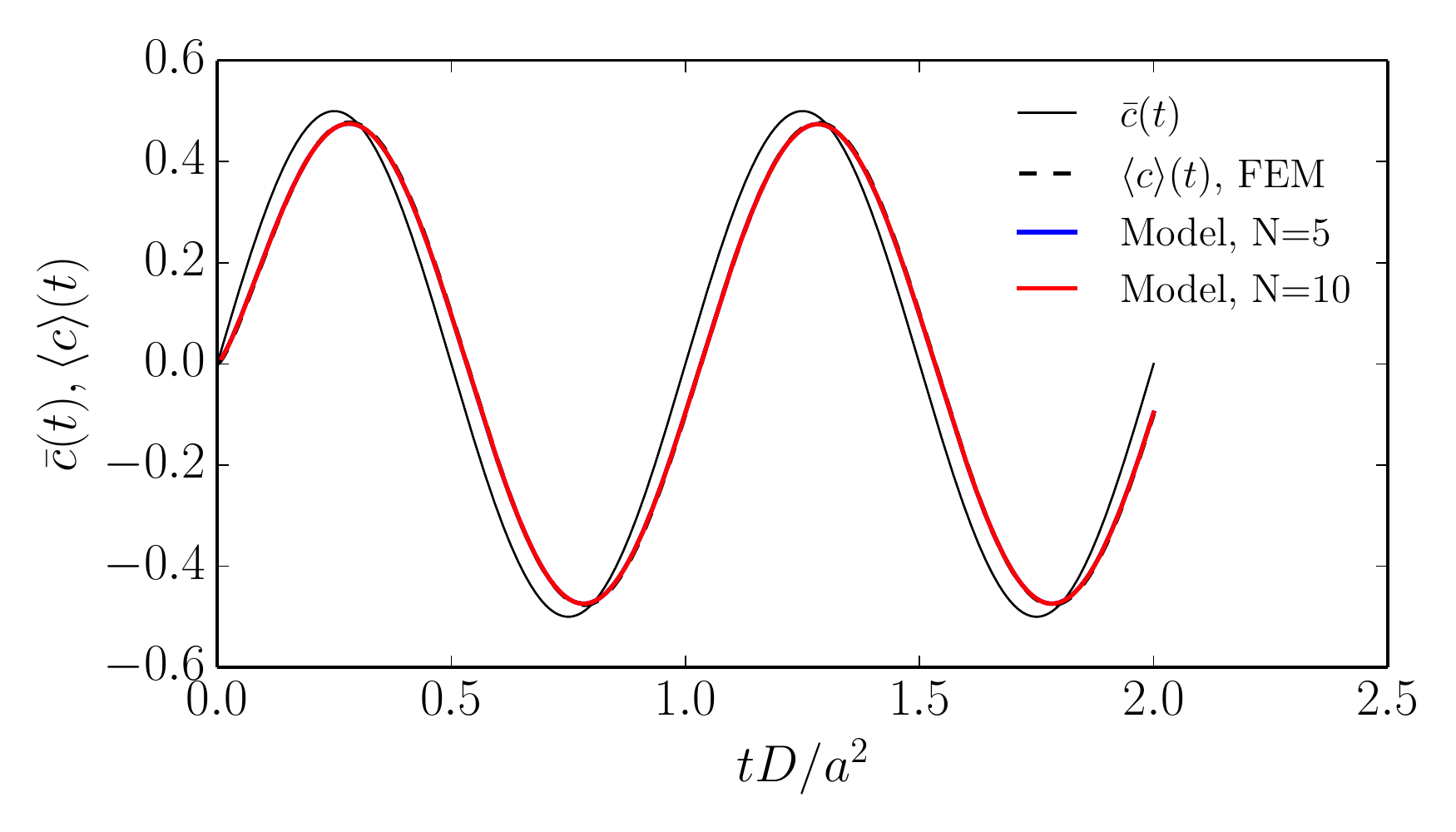}
\caption{}
\end{subfigure}
\begin{subfigure}{0.45\textwidth}
\includegraphics[width=\textwidth]{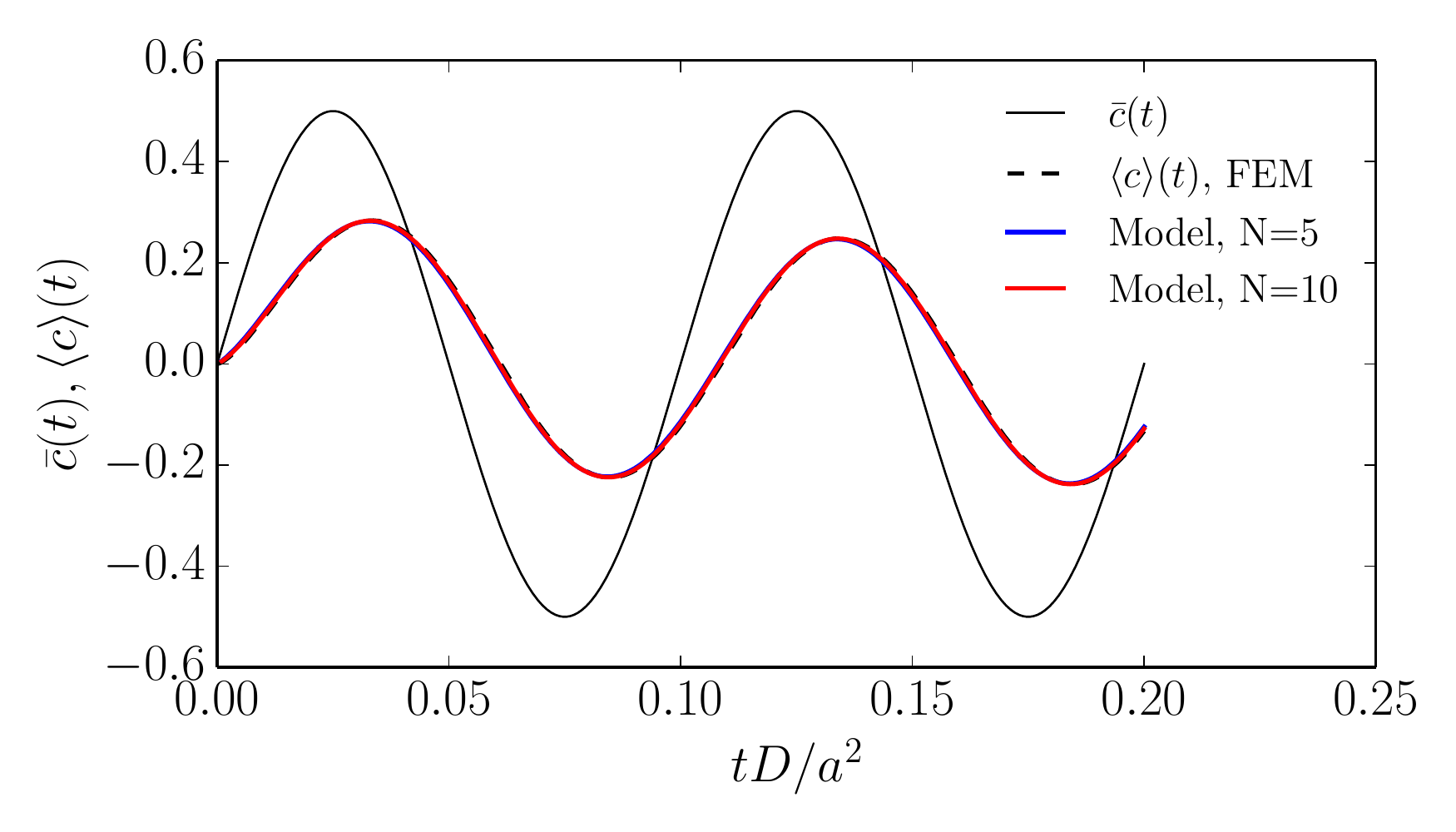}
\caption{}
\end{subfigure}
\begin{subfigure}{0.45\textwidth}
\includegraphics[width=\textwidth]{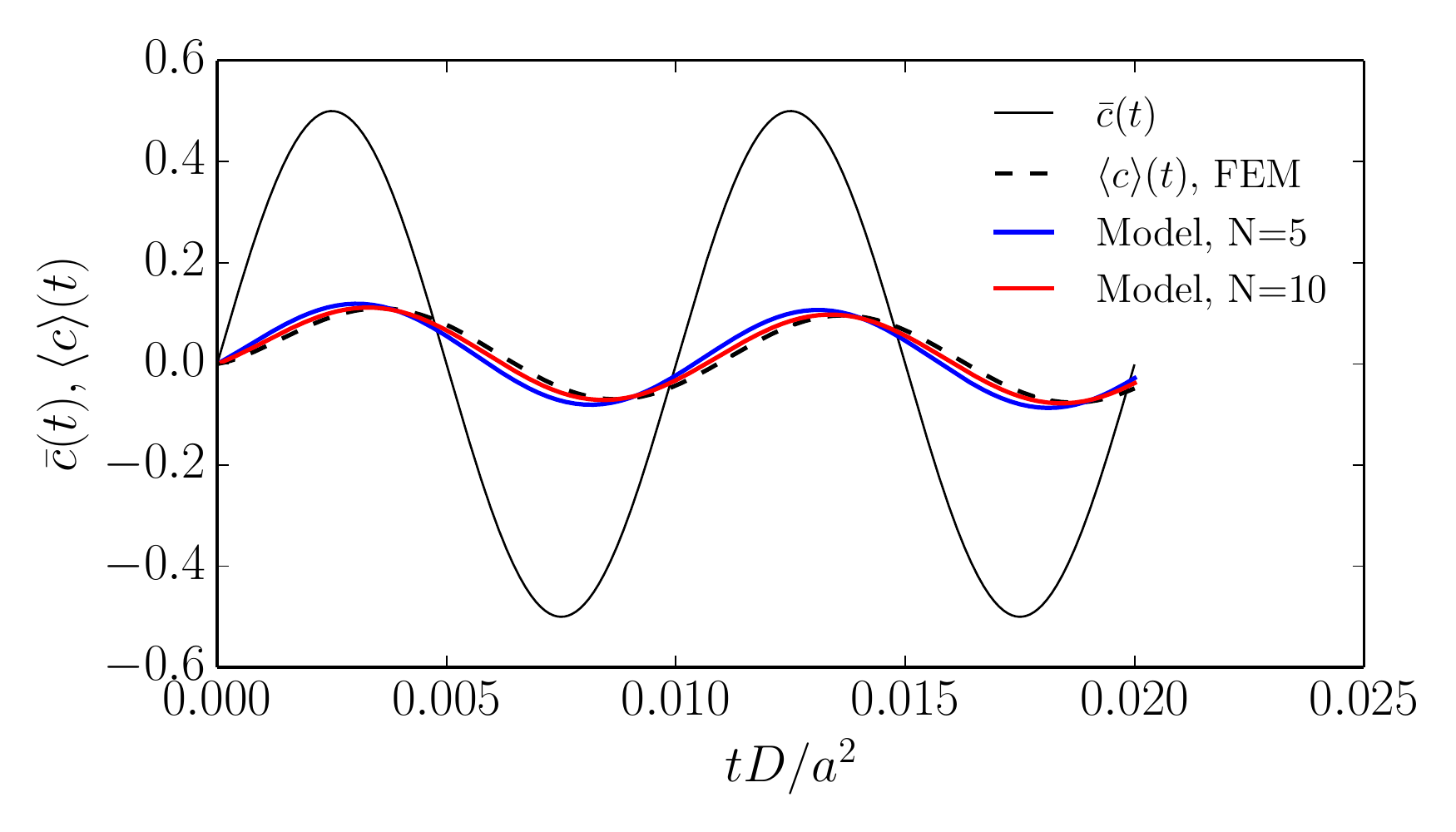}
\caption{}
\end{subfigure}
\caption{Evolution of the average concentration in a star-shaped domain subject to harmonic boundary condition $\bar c(t)$ of various periods $T$, with a) $T=10a^2/D$, b) $T=a^2/D$, c) $T=0.1a^2/D$ and d) $T=0.01a^2/D$. Reference results are provided by full-field FE simulations. Model predictions are based on the estimated creep function obtained from FE modal analysis.}
\label{fig:star_harmonic}
\end{center}
\end{figure}

Figure \ref{fig:star_harmonic} shows the effective response of the star-shaped inclusion subjected to sinusoidal boundary condition: $\bar c(t) = \sin(\omega t)$, with $\omega = 2\pi/T$ the angular frequency and $T$ the excitation period. Model predictions are shown for $N=5$ and $N=10$. As the frequency increases, the amplitude of the average concentration response decreases and a phase lag develops. The quality of the model predictions for increasing frequencies is directly correlated to the accuracy of the creep function at short time scales. For a large excitation period, the response is dominated by the long-time inclusion response, and a small number of modes is required for an accurate prediction  (Figures \ref{fig:star_harmonic}a-c). A higher number of modes is required to also capture the short-time response when the excitation period decreases (Figure \ref{fig:star_harmonic}d).  

\begin{figure}
\begin{center}
\begin{subfigure}{0.45\textwidth}
\includegraphics[width=\textwidth]{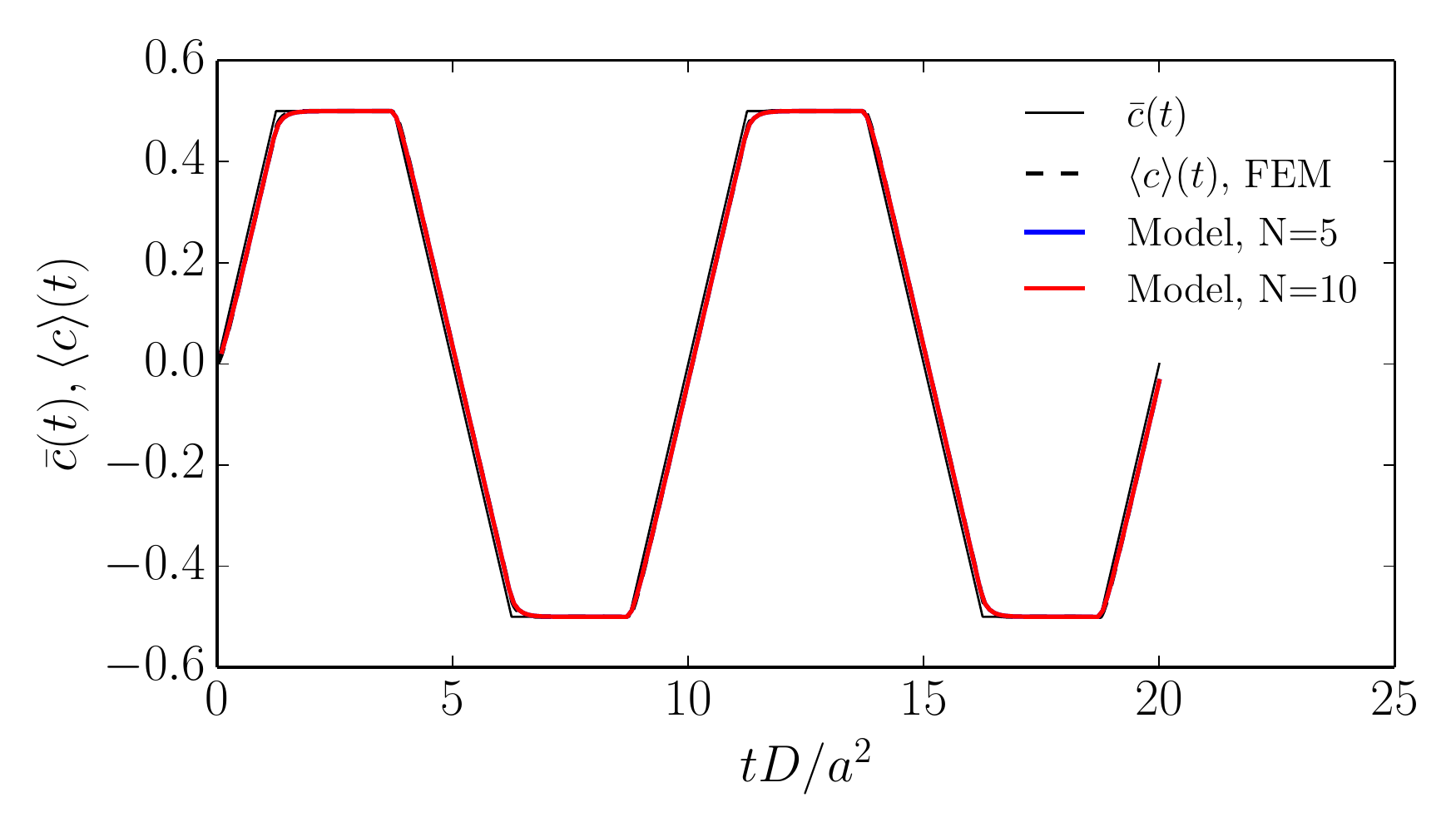}
\caption{}
\end{subfigure}
\begin{subfigure}{0.45\textwidth}
\includegraphics[width=\textwidth]{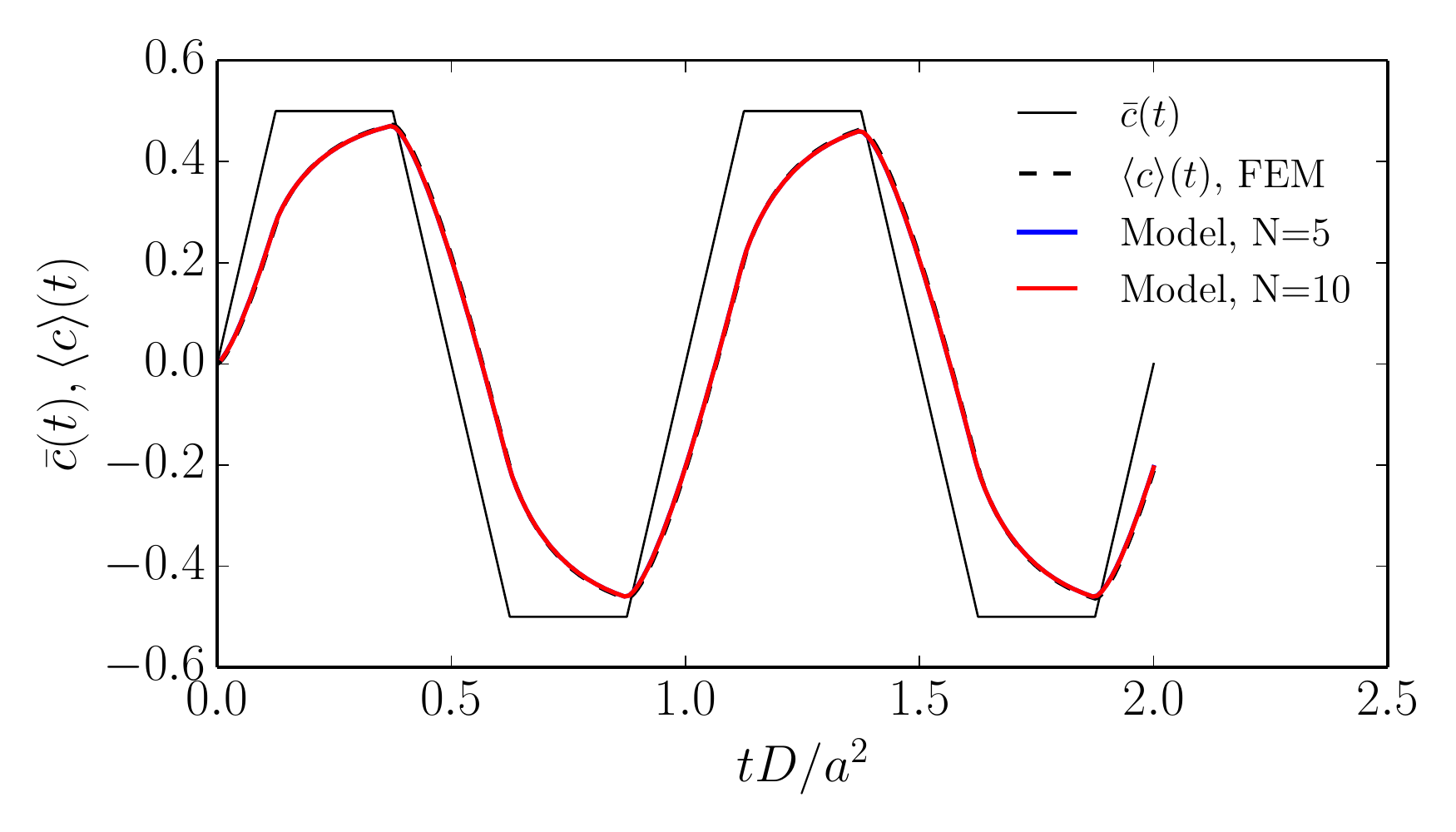}
\caption{}
\end{subfigure}
\begin{subfigure}{0.45\textwidth}
\includegraphics[width=\textwidth]{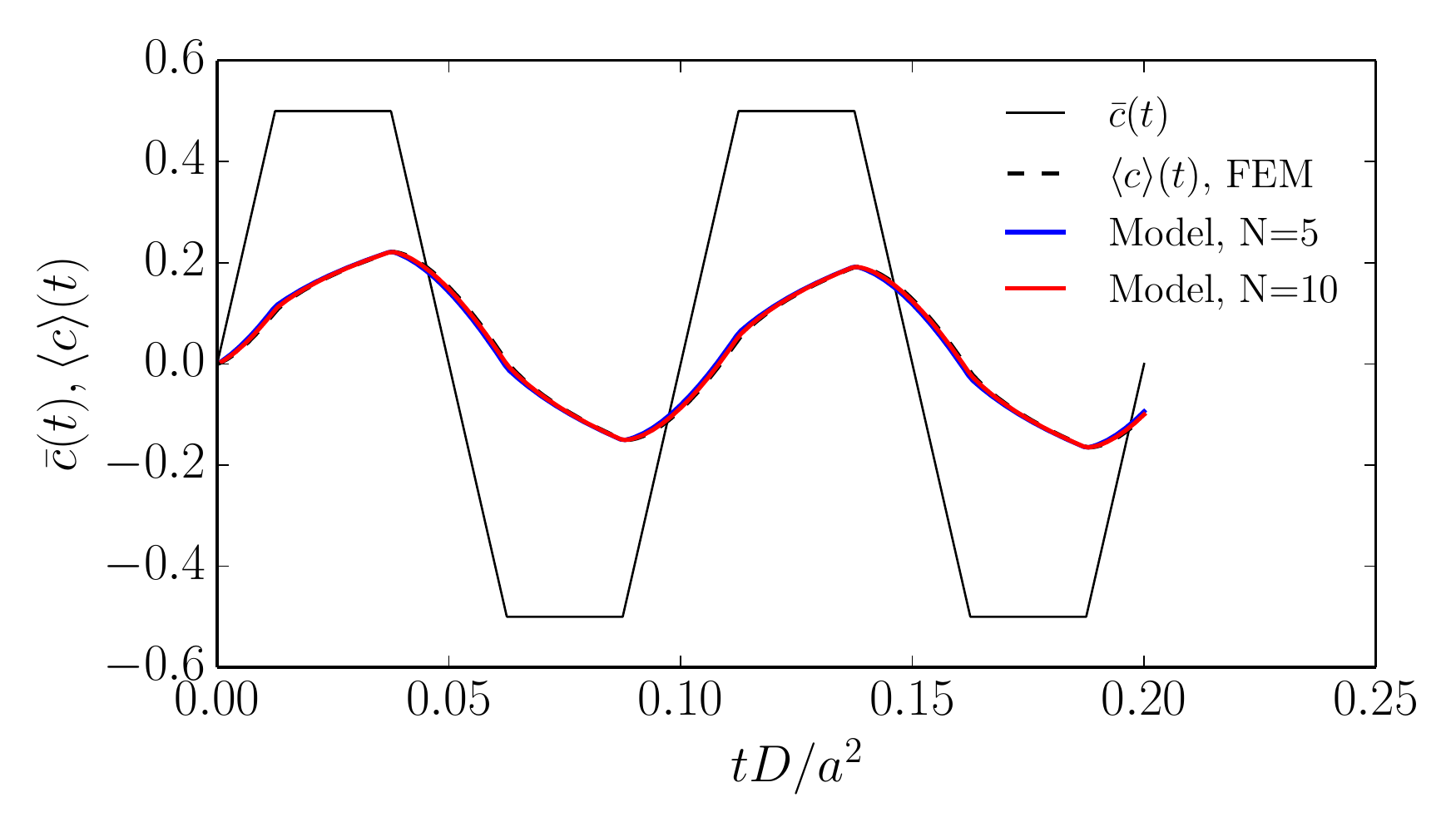}
\caption{}
\end{subfigure}
\begin{subfigure}{0.45\textwidth}
\includegraphics[width=\textwidth]{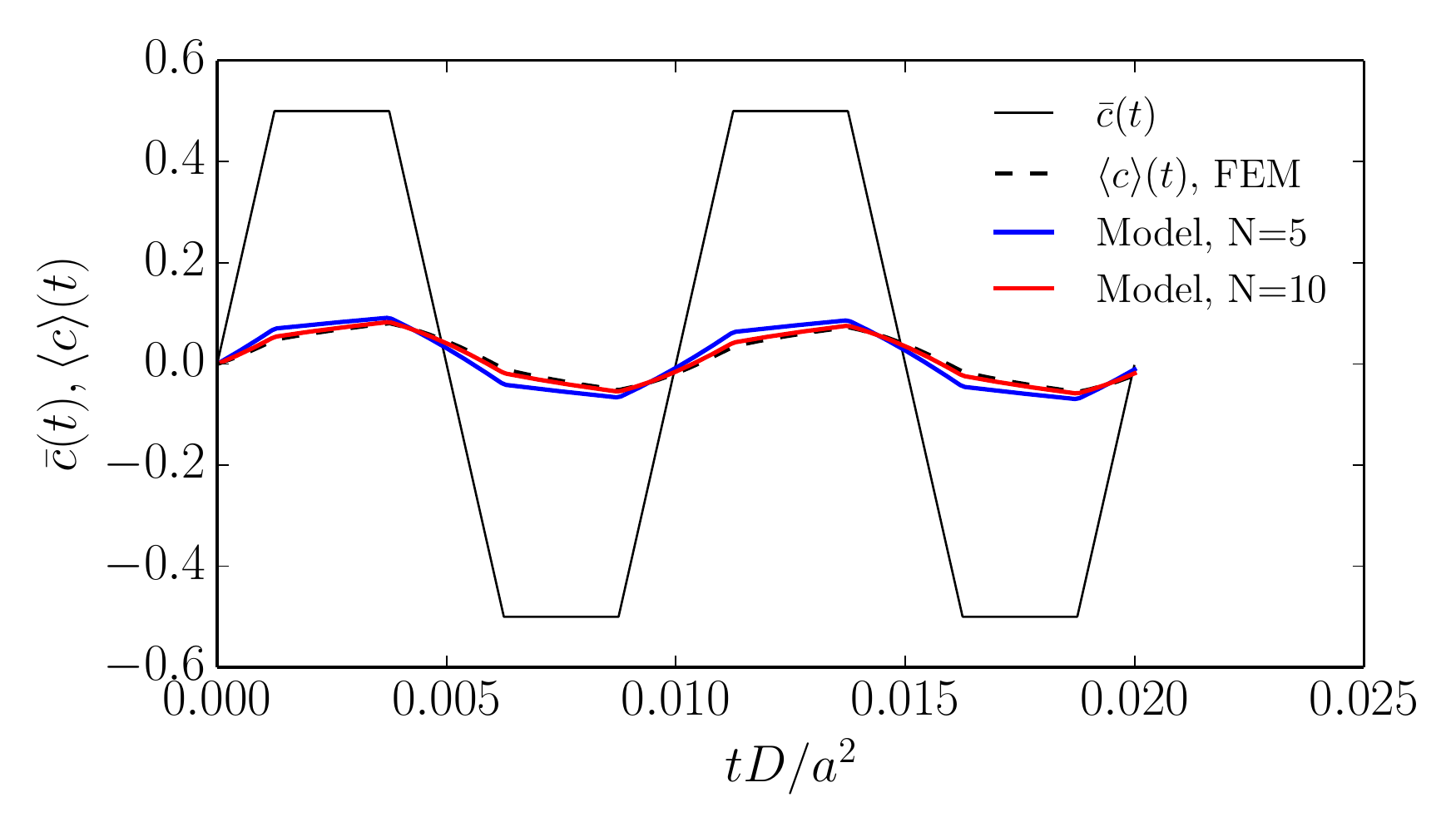}
\caption{}
\end{subfigure}
\caption{Evolution of the average concentration in a clover-shaped domain subject to cyclic boundary condition $\bar c(t)$ of various periods $T$, with a) $T=10a^2/D$, b) $T=a^2/D$, c) $T=0.1a^2/D$ and d) $T=0.01a^2/D$. Reference results are provided by full-field FE simulations. Model predictions are based on the estimated creep function obtained from FE modal analysis. }
\label{fig:clover_cyclic}
\end{center}
\end{figure}

As another example, Figure \ref{fig:clover_cyclic} shows the response of the clover-shaped inclusion subjected to a cyclic loading consisting of alternating sequences of ramps and plateaus, for different frequencies. Similar to the previous example, the accuracy of the semi-analytical predictions increases with the number of modes $N$, and decreases with the loading frequency for a given $N$. However, the predictions are still very good for $N=10$ and $T=0.01 a^2/D$.

\section{Mean-field estimates for nonlinear diffusion problems} \label{sec:nonlinear}
The estimates based on a finite number of internal variables presented in Sections \ref{sec:estimates} and \ref{sec:results} all rely on the assumption of constant diffusion coefficient, condition under which the exact series representation (\ref{eq:creep_fun}) of the chemical creep function holds. In this section we briefly address the more general case of concentration-dependent diffusion coefficient: $D=D(c)$. In this case, the general expression (\ref{eq:Fick_2ndlaw}) of Fick's second law needs to be used and the diffusion problem becomes nonlinear. The exact results presented in Section \ref{sec:exact} - in particular expression (\ref{eq:creep_fun}) for the chemical creep function - do not longer hold.

We propose a simple mean-field approach to generalise the estimates of Section \ref{sec:estimates} to the case of concentration-dependent diffusion coefficient. The approach is based on the observation that, in the linear diffusion problem, the relaxation times are each inversely proportional to the diffusivity. This property also holds for the estimated relaxation times obtained with FE modal analysis. Let $\tilde A_n^{lin}$ and $\tilde{\tau}^{lin}_n$ be these estimated coefficients and relaxation times in the linear case with constant diffusivity $D_0$. In the nonlinear case $D=D(c)$, we assume that $\langle c \rangle = \sum_{n=1}^{N+1} b_n$, where the internal variables $b_n$ still obey the differential system (\ref{eq:ODE_bn})-(\ref{eq:bnp1}):
\begin{eqnarray} 
\tilde{\tau}_n \dot{b}_n + b_n &=& \tilde A_n \bar c \label{eq:ODE_nonlinear1}\\
b_{N+1} &=& \tilde A_{N+1} \bar c \label{eq:ODE_nonlinear2}
\end{eqnarray}
with coefficients now given by: 
\begin{eqnarray} 
\tilde A_n &=& \tilde A^{lin}_n, \\
\tilde{\tau}_n &=& \tilde{\tau}^{lin} \frac{D_0}{D(\langle c \rangle)}.
\end{eqnarray}
In the mean-field model, the relaxation times are thus assumed to depend on the average concentration through the concentration-dependency of the diffusion coefficient. This assumption introduces a coupling between the ODEs (\ref{eq:ODE_nonlinear1}). In practice, the system (\ref{eq:ODE_nonlinear1})-(\ref{eq:ODE_nonlinear2}) can still be solved in a full-implicit way, but now requires the solution of a linear algebraic system at each time step. The effective inclusion response (\ref{eq:eff_rel}) is now of the form:
\begin{equation}
\frac{d\langle c \rangle}{dt} = \sum_{n=1}^N \frac{\tilde A_n \bar c - b_n}{\tau_n(\langle c \rangle)} + \tilde A_{N+1} \frac{d\bar c}{dt}.
\end{equation}

\begin{figure}
\begin{center}
\includegraphics[width=0.5\textwidth]{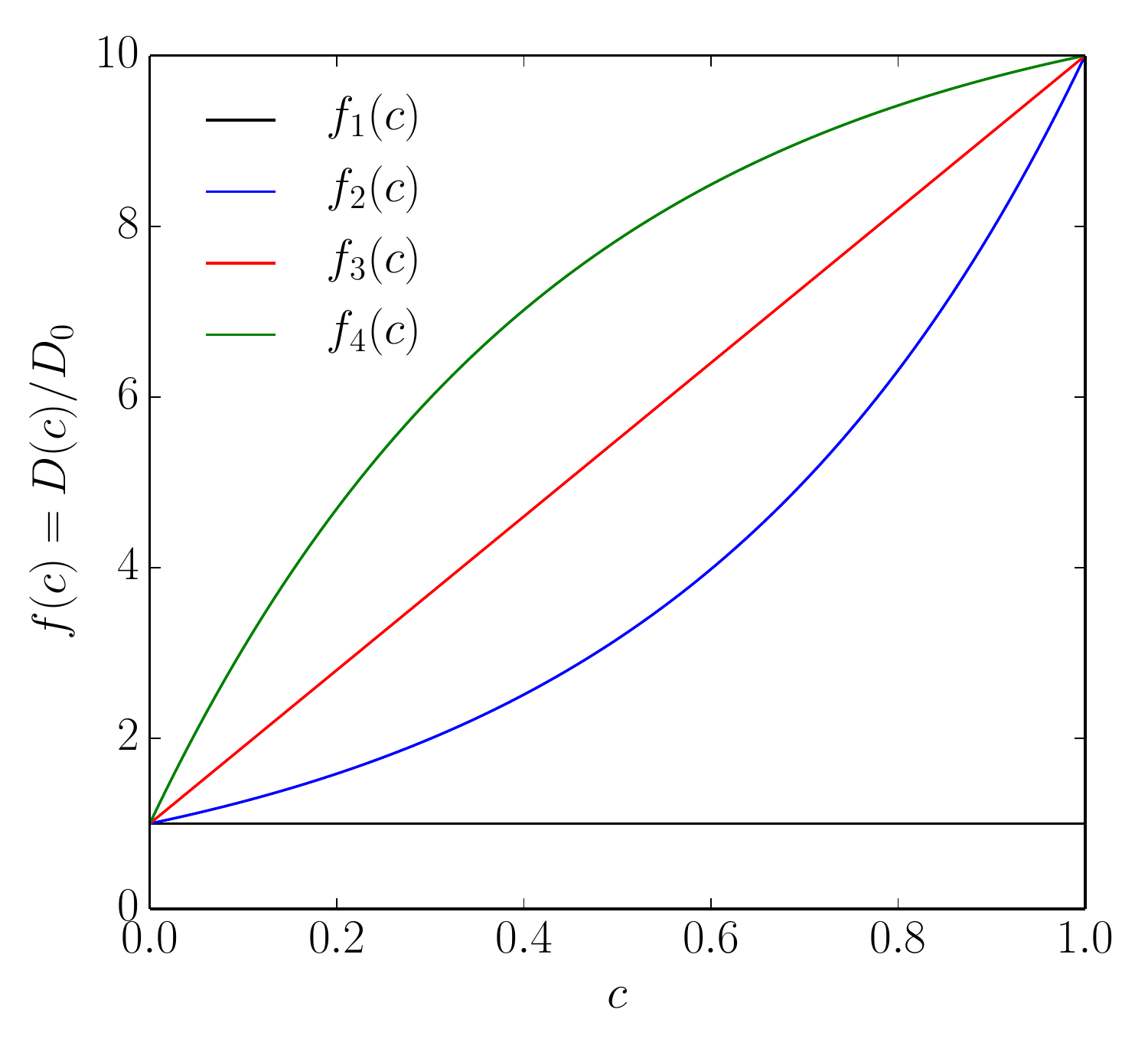}
\caption{The normalised concentration-dependent diffusion coefficient $D/D_0 = f(c)$. The four different functions $f_1(c)$, $f_2(c)$, $f_3(c)$ and $f_4(c)$ are reported in Equations (\ref{f1})-(\ref{f4}).}
\label{fig:Dfun}
\end{center}
\end{figure}

We illustrate the proposed approach by considering diffusion coefficients of the following form: 
\begin{equation}\label{eq:nonlinear_D}
D(c) = D_0 f(c),
\end{equation}
where $D_0$ is the constant reference value used to identify the coefficients $\tilde A_n^{lin}$ and relaxation times $\tilde{\tau}^{lin}_n$ of the comparison linear problem, and $f(c)$ a dimensionless function of concentration. The following functions are used \cite{crank1975}:
\begin{eqnarray}
f_1(c) &=& 1 \label{f1}\\
f_2(c) &=& \exp(2.303c) \label{f2}\\
f_3(c) &=& 1 + 9c \label{f3}\\
f_4(c) &=& 1 + 10(1-\exp(-2.303c)) \label{f4}
\end{eqnarray}
These functions are illustrated in Figure \ref{fig:Dfun}. 

We illustrate the mean-field approach in the case of a cylindrical inclusion or radius $a$ subjected to a harmonic loading: $\bar c(t) = 0.5 + 0.5 \sin(\omega t - \pi/2)$. This loading ensures that the concentration within the inclusion is always in the interval $[0,1]$. Here, $\omega = 2\pi/T$ is the angular frequency, and $T$ the period. Reference solutions are obtained by solving the nonlinear diffusion problem in the cylinder by the FE method. Accounting for axisymmetry, it reduces to a 1D problem along the radius, and we used 100 equispaced nodes and linear shape functions. The coefficients $\tilde A_n^{lin}$ and relaxation times $\tilde{\tau}^{lin}_n$ of the linear comparison problem are obtained from their analytical expressions for the linear diffusion in a cylinder (Table \ref{tab:table1}), and we used the first 10 modes. This number of modes is sufficiently high so that the discrepancy between the reference and mean-field estimate come solely from the nonlinearity, and not from the error on the estimate in the linear diffusion problem. 

\begin{figure}
\begin{center}
\begin{subfigure}{0.6\textwidth}
\includegraphics[width=\textwidth]{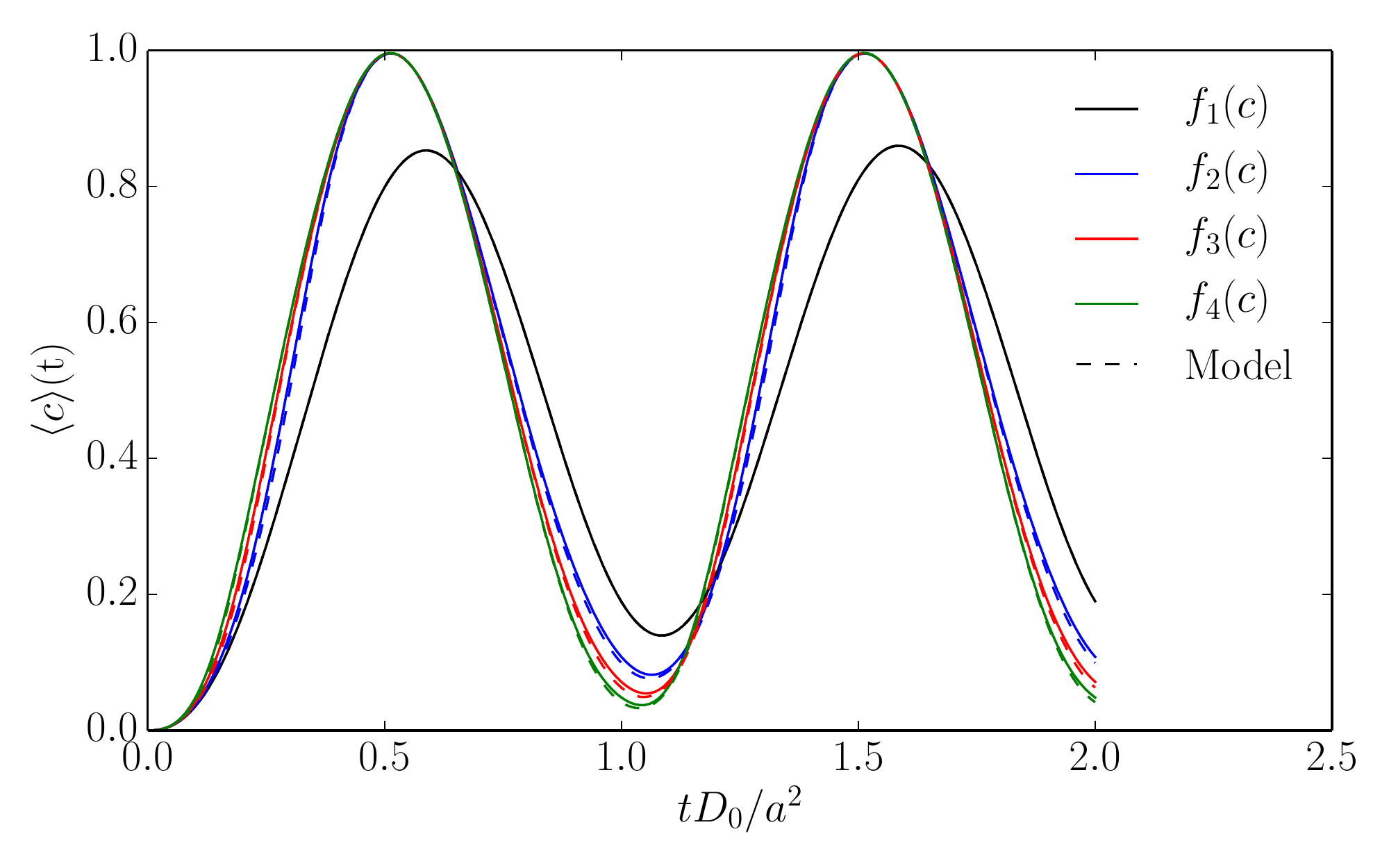}
\caption{}
\end{subfigure}
\begin{subfigure}{0.6\textwidth}
\includegraphics[width=\textwidth]{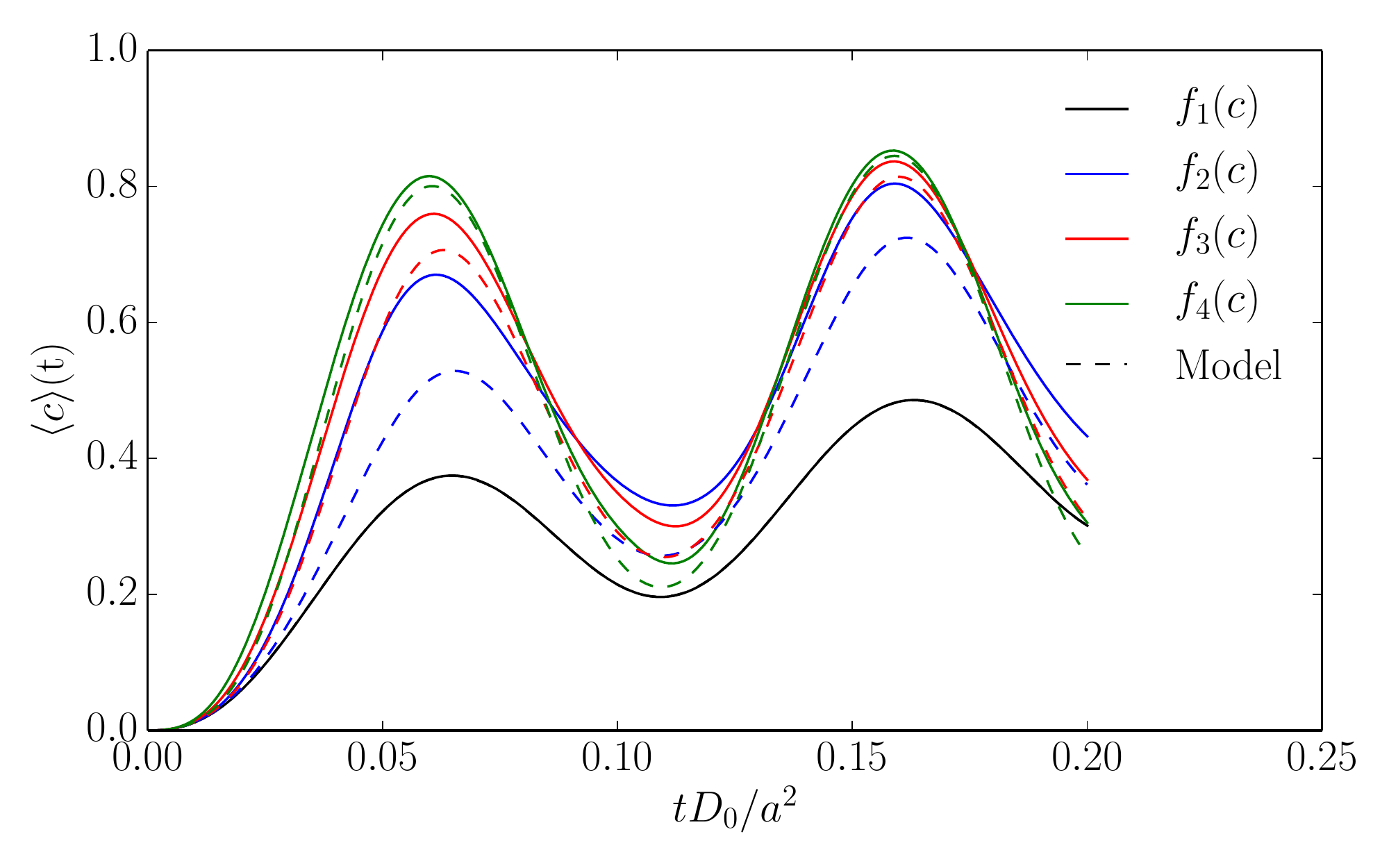}
\caption{}
\end{subfigure}
\caption{Mean-field solution to the radial diffusion problem in a cylinder subjected to harmonic excitation with period a) $T=a^2/D$ and b) $T=0.1a^2/D$. The diffusion coefficient varies with concentration, $D(c)=D_0 f(c)$. The continuous lines are the reference FE predictions, and the dashed lines are the mean-field estimates. For the latter, the first 10 coefficients $A_n$ and relaxation times $\tau_n$ of the analytical solution for the cylinder (Table \ref{tab:table1}) were used.}
\label{fig:nonlin}
\end{center}
\end{figure}

The comparison between the full-field reference predictions and the mean-field estimate is shown in Figure \ref{fig:nonlin}. The quality of the predictions is highly sensitive to the frequency of the applied load, as well as to the particular concentration dependency of the diffusion coefficient. At low frequency ($T=a^2/D$, Figure \ref{fig:nonlin}a), the mean-field predictions are in very good agreement with the reference FE solution. However, at higher frequency ($T=0.1a^2/D$, Figure \ref{fig:nonlin}b), some significant discrepancies are found for $f(c)=f_2(c)$ and $f(c)=f_3(c)$, while the model predictions for $f(c)=f_4(c)$ are still acceptable. For both frequencies, one verifies that application of the model to the case of constant coefficient ($f(c)=1$) with $N=10$ gives results which are indistinguishable from the reference solution, which confirms that the error in nonlinear cases stem from the non-linearity only.

\section{Conclusion}
In this work we have developed semi-analytical estimates for the effective transient response of inclusions subjected to uniform, time-varying Dirichlet boundary conditions. The proposed estimates are of the general form (\ref{eq:eff_rel}) or (\ref{eq:eff_rel2}) and involve a finite number of internal variables accounting for the loading history dependency of the behaviour. The proposed estimates admit simple representations in terms of equivalent spring-dashpot rheological models, and can readily be integrated in time using a fully implicit scheme. The estimates were constructed based on the chemical creep function of the inclusion, to be identified once and for all by solving the inclusion problem under a step loading. While linear diffusion problems have been studied for a long time, we believe that our presentation based on the notion of chemical creep function and the analogy with viscoelasticity is original.

Here we have proposed three methods to construct semi-analytical estimates, namely a collocation method, a constrained optimisation method, and a FE-based modal analysis approach. The first two methods directly inspire from existing methods in viscoelasticity. The third method constitutes an original application of FE-based modal analysis for estimating the effective transient diffusion behaviour. We have shown that these estimates correctly capture the effective transient response of inclusions under time-varying loading conditions with a very small number of relaxation modes, or equivalently internal variables, at only a fraction of the cost of full-field simulations. We also proposed a simple extension to the case of nonlinear diffusion problems. We hypothesise that other - possibly enhanced -  estimates could be developed in both the linear and nonlinear cases  based on the idea of chemical creep function.

The proposed estimates for the effective transient response of inclusions can be used as the building blocks of more sophisticated multi-inclusion models, for example to investigate the role of inclusion shape and size distribution on the memory effect. In this case, it is necessary to consider a distribution of inclusions in a matrix, and to apply macroscopic loading conditions on the boundary of a RVE of the microstructure. The chemical loading history on the boundary of a particular inclusion in the RVE then becomes dependent on the matrix chemical properties and RVE geometry. In a mean-field approach, one needs to propose a suitable localisation rule in order to determine the chemical load experienced by each inclusion in the RVE at a given time. For example, under the assumption of an infinitely fast diffusing matrix, the chemical loading can be expressed in terms of an effective chemical potential assumed uniform over each inclusion boundary, but varying with the inclusion position. One can then use expression (\ref{eq:eff_rel2}) for each inclusion, and the total, effective inclusion response is obtained by averaging over all the inclusions. These developments are out of scope of the present paper. A comprehensive homogenisation framework for transient diffusion problems in composites, as well as mean-field estimates based on the present work, will be presented in a forthcoming contribution \cite{brassart2018}.

\section*{Acknowledgments}
The authors thank Thomas Heuz\'e for insightful discussions on modal methods in transient heat transfer.

\begin{appendices}
\renewcommand*{\thesection}{\appendixname~\Alph{section}}
\section{Regime response of inclusions subjected to harmonic loading}
We address the regime response of an inclusion subjected to harmonic excitation of the form $\bar c e^{i\omega t}$on its boundary, where $\omega$ is the angular frequency and $\bar c$ a constant real number. In regime, local solutions are of the form $\phi(\bm x)e^{i \omega t}$, with $\phi(\bm x)$ a complex function of the spatial coordinates which satisfies:  
\begin{equation} \label{eq:PDE_harmonic}
i\omega \phi = D \bm{\nabla}^2 \phi.
\end{equation}
The average concentration is then: $\langle c \rangle =  \langle \phi \rangle e^{i\omega t}$. 

In analogy with linear viscoelasticity, we introduce the complex chemical creep modulus $J^*$, its "storage" and "loss" components $J'$ and $J''$, and a loss tangent:
\begin{equation} \label{eq:complex_mod}
J^* = \frac{\langle \phi \rangle}{\bar c} = J' + i J'',\quad \tan \delta = \frac{J''}{J'}.
\end{equation}
The storage and loss moduli respectively represent the in-phase and out-of-phase components of the inclusion response, relative to the applied concentration. In practice, we are interested in the real part of the average concentration response: 
\begin{equation}\label{eq:harmonic_real_part}
Re\left[ \langle \phi \rangle e^{i\omega t} \right] = \frac{\bar c J'}{\cos \delta} \cos (\omega t + \delta).
\end{equation}    
This expression identified the amplitude of the average concentration as $\bar c J'/\cos \delta$. 

We relate the storage and loss moduli to the chemical creep function (\ref{eq:creep_fun_def}) by following a procedure previously proposed for linear viscoelasticity, see \cite{christensen1982}, Section 1.6. First introduce the harmonic boundary condition $\bar c e^{i\omega t}$ in the integral representation (\ref{eq:integral_representation2}) of the inclusion response: 
\begin{equation}
\langle \phi \rangle e^{i \omega t} = \bar c \int_{-\infty}^{t} J(t-t') i\omega e^{i\omega t'} dt',  
\end{equation}
where we have used $t=-\infty$ as the lower integration bound to describe an infinitely cycling behaviour. Next we introduce the change of variable $\eta = t-t'$ to rewrite the above equation as:
\begin{equation}
\langle \phi \rangle = i\omega \bar c \int_{-\infty}^{\infty} H(\eta) J(\eta) e^{-i\omega\eta} d\eta .
\end{equation}
Recognising the definition of the Fourier transform of the creep function in the right-hand side of the last equation, we write:
\begin{equation} \label{eq:complex_mod_series}
J^* = \sum_{n=1}^{\infty} A_n \frac{1}{1+i\omega \tau_n},
\end{equation}
where we also used the identity (\ref{eq:sum_An}). From the series representation (\ref{eq:complex_mod_series}), the real and imaginary parts can be explicitly obtained:
\begin{eqnarray}
J' &=& \sum_{n=1}^{\infty}A_n \frac{1}{1+ (\omega \tau_n)^2} ,\label{eq:storage}\\
J'' &=& - \sum_{n=1}^{\infty} A_n \frac{\omega \tau_n}{1 + (\omega \tau_n)^2 }. {\label{eq:loss}}
\end{eqnarray}
While the creep function (\ref{eq:creep_fun}) is given by an infinite series of exponentials, the complex modulus is given by an infinite series of complex fractions. Both series involves the same coefficients $A_n$ and $\tau_n$.  

Alternatively, the complex modulus can be obtained in closed form in the case of the 1D linear diffusion problems in a plane sheet, a cylinder or a sphere. The PDE (\ref{eq:PDE_harmonic}) then reduces to one single ODE that can be solved analytically for the function $\phi(r)$, with $r$ the spatial coordinate \cite{carslaw1959}. We list the final results for both the local solution and the average response in Table \ref{tab:table3}. In the table, $a$ is a characteristic size (cf. Figure \ref{fig:creepfun}), $k$ is the wavenumber, $k=(\omega/D)^{1/2}$, and $I_0$ and $I_1$ are modified Bessel functions. We have verified numerically that the real and imaginary parts of the closed form expressions for $\langle \phi \rangle/\bar c$ reported in Table \ref{tab:table3} and the series expressions for $J'$ and $J''$ coincide.

\begin{figure}
\begin{subfigure}{0.45\textwidth}
\includegraphics[width=\textwidth]{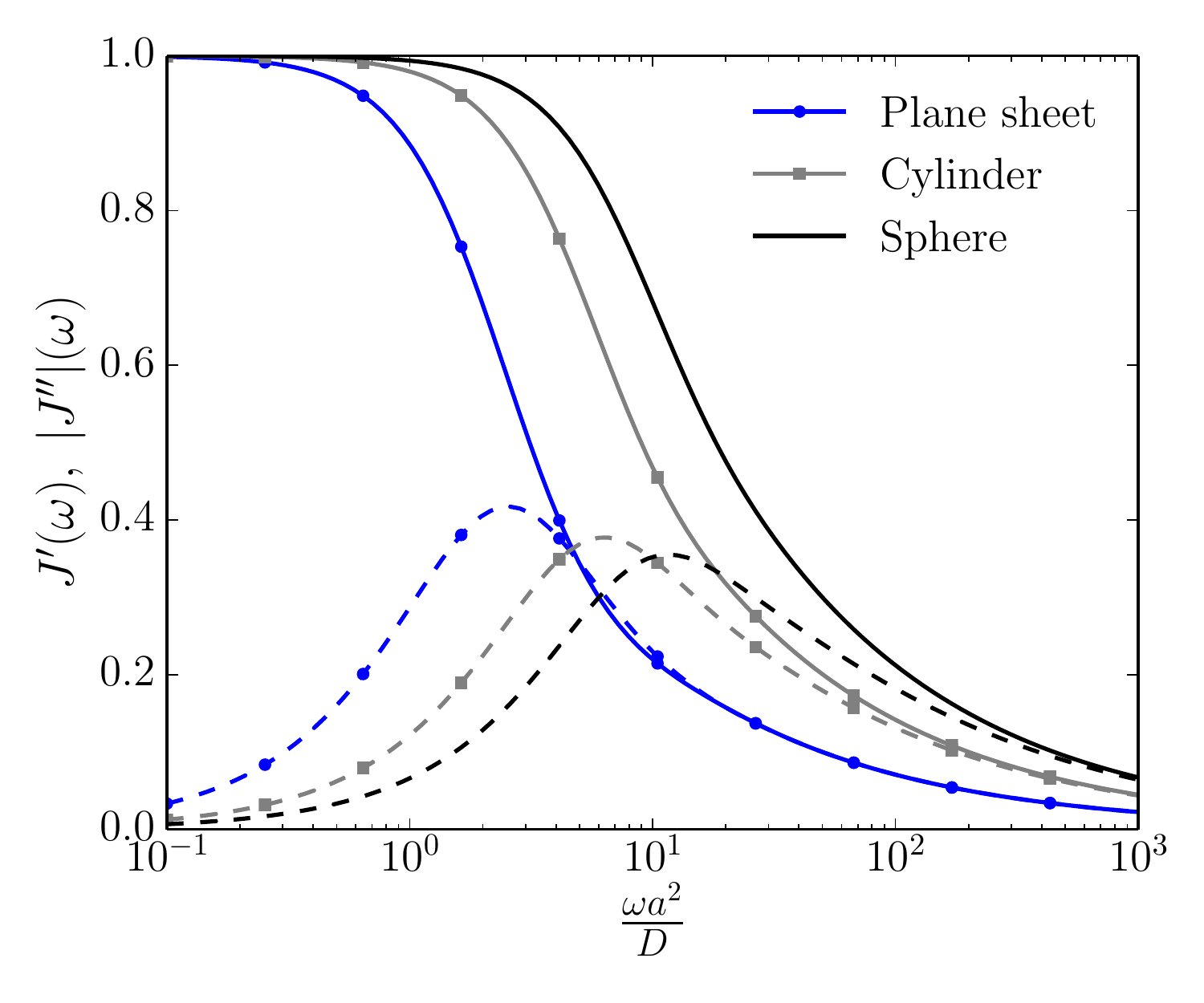}
\caption{}
\end{subfigure}
\begin{subfigure}{0.45\textwidth}
\includegraphics[width=\textwidth]{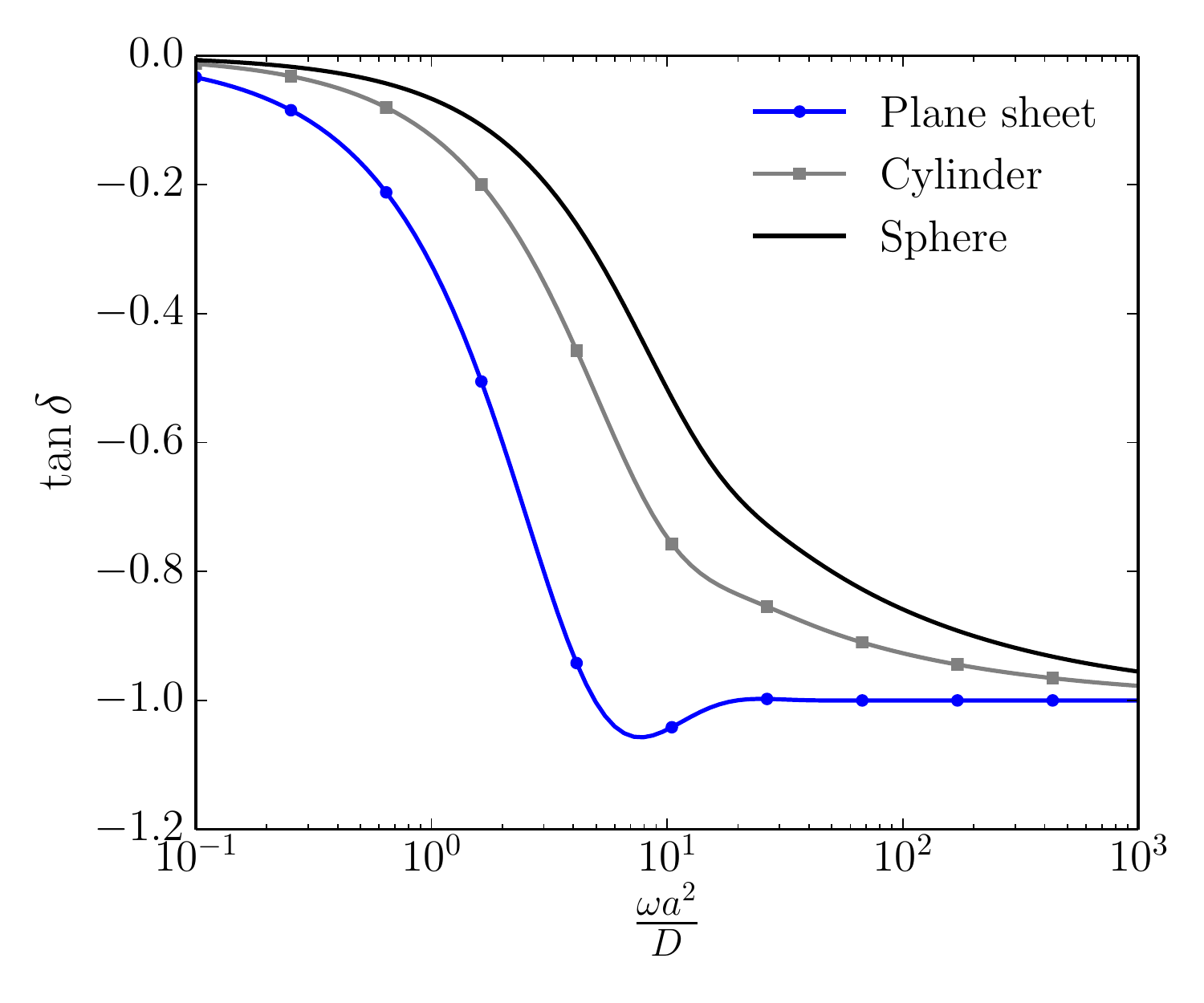}
\caption{}
\end{subfigure}
\caption{Frequency response. (a) Real and Imaginary parts of the complex relaxation modulus and b) the loss tangent.}
\label{fig:relax_freq}
\end{figure}

The evolution of the storage and loss moduli with the frequency of the applied chemical load is represented in Figure \ref{fig:relax_freq}a for the planar, cylindrical and spherical geometries. Similar curves were previously derived by Auriault \cite{auriault1983}. The storage modulus tends to $J'=1$ at low frequencies (chemical equilibrium), and to $J'=0$ at large frequencies. Different from viscoelasticity, there is a vanishing instantaneous concentration response when the applied load is very fast. The loss modulus tends to zero for both low and high frequencies, and displays a peak at intermediate frequencies, similar to viscoelasticity. However, the loss modulus remains below the storage modulus at all frequencies (in absolute value) in the case of a cylindrical and spherical geometry, and slightly exceeds the storage modulus in a small range of frequencies in the case of the planar geometry. This is also different from classical viscoelasticity where the loss modulus may exceed the storage modulus by a significant amount and for a significant range of frequencies. Noticeably, the storage and loss moduli converge to each other at high frequencies.       

The loss tangent is represented in Figure \ref{fig:relax_freq}b. At low frequencies, the average response is in phase with the applied loading (chemical equilibrium), $\delta = 0$. At high frequencies, the loss tangent tends to $-1$, which corresponds to a phase angle $\delta = \frac{\pi}{4}$. In other words, the concentration lags behind the applied surface concentration with a maximum of $\pi/4$ phase lag at high frequencies. The loss tangent monotonically decreases from 0 to -1 for the circular geometries. In contrast, values below -1 are reached for intermediate frequencies in the case of the planar geometry, and this occurs in the frequency range where the loss modulus exceeds (in absolute value) the storage modulus.

\begin{table}[h]
\centering
\caption{Solution of harmonic diffusion problem in a plane sheet of thickness $2a$, a cylinder and a sphere of radius $a$. $r$ is the spatial coordinate, and $k$ is the wavenumber, $k=(\omega/D)^{1/2}$.}
\label{tab:table3}
\begin{tabular}{lccc}
\toprule
& Plane sheet & Cylinder & Sphere \\
\midrule
$ \frac{\phi(r)}{\bar c} $ & $ \frac{\cosh\left( k(a-r)i^{1/2}\right)}{\cosh \left( kai^{1/2} \right)} $ & $\frac{I_0 \left( kri^{1/2} \right)}{I_0 \left( kai^{1/2} \right) }$ &$ \frac{a}{r} \frac{\sinh\left( k r i^{1/2} \right)}{\sinh\left(  kai^{1/2} \right)}$ \\
$ \frac{\langle \phi \rangle}{\bar c} $ & $ \frac{\tanh\left( kai^{1/2} \right)}{kai^{1/2}} $ & $ 1 + \frac{J_2 \left( ka(-i)^{1/2} \right) }{J_0 \left( ka(-i)^{1/2} \right) } $ & $\frac{3}{kai^{1/2}} \frac{\cosh\left( kai^{1/2} \right)}{\sinh \left( k a i^{1/2} \right)} - \frac{3}{(ka)^2 i } $ \\
\bottomrule
\end{tabular}
\end{table}

\end{appendices}

\end{document}